\documentclass[11pt, a4paper]{article}
\pdfoutput=1
\usepackage[utf8]{inputenc}
\usepackage{amsmath}
\usepackage{mathtools}
\usepackage{braket}
\usepackage{bbold}
\usepackage{amsfonts}
\usepackage{amssymb}
\usepackage{geometry}
\usepackage{csquotes}
\usepackage{slashed}
\usepackage{tensor}
\usepackage{fancyhdr}
\usepackage{wrapfig}
\usepackage[toc,page]{appendix}
\usepackage[english]{babel}
\usepackage{hyperref}
\usepackage{graphicx}
\usepackage{multicol}
\usepackage{lipsum}
\usepackage{mwe}
\usepackage[export]{adjustbox}
\usepackage{xcolor}
\usepackage[skip=5pt]{caption}
\usepackage{subcaption}
\usepackage{placeins}
\usepackage{authblk}
\usepackage[
backend=biber,sorting=none,doi=false,isbn=false,url=false,maxnames=10
]{biblatex}
\DeclareMathOperator{\Tr}{Tr}
\graphicspath{ {./p_wave_figs/} }
\begin{document}

\begin{flushright}
QMUL-PH-22-31\\
\end{flushright}

\bigskip
\bigskip 
\bigskip

\centerline{\Large What lies beyond the horizon of a holographic p-wave superconductor}

\bigskip
\bigskip

\centerline{\bf Lewis Sword and David Vegh}

\bigskip

\begin{center}

\small{
{ \it Centre for Theoretical Physics, Department of Physics and Astronomy \\
Queen Mary University of London, 327 Mile End Road, London E1 4NS, UK}}

\medskip

{\it Email:} \texttt{l.sword@qmul.ac.uk}, \texttt{d.vegh@qmul.ac.uk}

\end{center}

\bigskip 
\bigskip
\centerline{ \it \today}

\begin{abstract}
We study the planar anti-de Sitter black hole in the p-wave holographic superconductor model. We identify a critical coupling value which determines the type of phase transition. Beyond the horizon, at specific temperatures flat spacetime emerges. Numerical analysis close to these temperatures demonstrates the appearance of a large number of alternating Kasner epochs.

\end{abstract}

\tableofcontents


\newpage
\section{Introduction}

The AdS/CFT correspondence \cite{Maldacena:1997re,Witten:1998qj,Gubser:1998bc}, otherwise known as gauge-gravity duality, introduced a method to inspect strongly coupled theories. From it emerged a dictionary relating fields in bulk spacetime to operators on its boundary. Using this correspondence, the gravitational dual of a superconductor, known as a holographic superconductor, was discovered \cite{Gubser:2008px,Hartnoll:2008vx,Hartnoll:2008kx}. By spontaneously breaking an abelian gauge symmetry of a charged scalar field in Schwarzschild-AdS spacetime, one produces a non-trivial expectation value for the scalar field, which corresponds to a non-zero condensate developing on the boundary\footnote{On the boundary the $U(1)$ symmetry that is spontaneously broken in the superconducting
phase is a global symmetry, hence is more accurately described as a superfluid.}. These original models used a simple $U(1)$ charged scalar boson to introduce a condensate (scalar ``hair"), however other bosonic condensate models are possible such as p-wave superconductors where a charged vector field is employed utilising $SU(2)$ gauge theory\footnote{In addition to the scalar and vector condensates, d-wave superconductors based on spin-2 condensates have also been the subject of similar holographic analysis \cite{Kim:2013oba}.}. This takes a $U(1)$ subgroup as electromagnetism and allows the gauge bosons that are charged under the $U(1)$ to condense. The original p-wave model was presented in \cite{Gubser:2008wv} followed by a top-down, string theory approach in \cite{Ammon:2008fc,Basu:2008bh,Ammon:2009fe} and backreaction on the metric was accounted for in \cite{Ammon:2009xh,Arias:2012py}.

In both vector and scalar condensate cases, the identification of the gravitational counterparts of superconductors is attributed to a condensing field and a black hole spacetime with a given Hawking temperature in the bulk. This topic has garnered great interest since its inception.
 The exploration of the black hole interior began with \cite{Frenkel:2020ysx,Hartnoll:2020rwq} and was extended to the full scalar field holographic superconductor model in \cite{Hartnoll:2020fhc}. Numerous interesting phenomena were observed in the interior including Kasner geometries\footnote{In the context of cosmology, Kasner regimes have also been investigated in \cite{Misner:1969hg,Belinsky:1970ew}.}, collapse of the Einstein-Rosen (ER) bridge and Josephson oscillations. The interior has subsequently been subjected to further study. This includes the introduction of additional field content and variation of coupling parameters \cite{Sword:2021pfm,Mansoori:2021wxf,Mirjalali:2022wrg,Liu:2021hap,Henneaux:2022ijt}, use of alternative black hole solutions \cite{Dias:2021afz,Grandi:2021ajl,Liu:2022rsy}, analysis of RG flows \cite{Caceres:2022smh,Caceres:2022hei}, as well as the construction of ``no inner-horizon" theorems \cite{Cai:2020wrp}. Investigation of the interior solutions for the p-wave model are now also being explored, with the analogous changes in geometry and matter fields being observed \cite{Cai:2021obq}. 

This paper aims to show the interesting changes in interior geometry by exploring the parameter space of the p-wave superconductor. The key result is that for a special selection of parameters, the numerical solutions appear to imply that the interior geometry becomes flat. Not only that, but either side of this parameter selection, the geometry becomes almost oscillatory in Kasner universes.

The outline of the paper is as follows. Section \ref{s:model} introduces the holographic p-wave superconductor model. Here the equations of motion are established, followed by details of the numerical procedure to solve them. We also state the equations' scaling symmetries, as well as the horizon and UV series expansions as governed by the necessary boundary conditions. Section \ref{s:ext_behaviour} puts the numerical solutions to use, by exploring the exterior of the black hole. We first analyse the field content in the exterior confirming that the solutions satisfy the correct boundary conditions. Under vanishing condensate, we enter normal phase as verified by the metric returning to that of a Reissner-Nordstr\"{o}m spacetime. Following this, upon various choices of the model parameters, the phase diagrams for the holographic superconductor are produced. The phase curves imply that a critical $g_{\text{YM}}$ coupling exists, which differentiates between first and second order transitions. This is confirmed by analysis of the grand potential derived from the Euclidean action as well as the entropy. Section \ref{s:int_behaviour} explores the interior and presents our main findings. We begin by studying the interior field content close to the horizon, which demonstrates typical behaviour previously seen such as the Josephson oscillations and ER bridge collapse. The focus then turns to specific points in the parameter space. Here the key finding is that at certain values the interior geometry appears to become flat while slight deviations away from this point lead to a highly oscillatory geometry comprised of individual Kasner universes for a given bulk radius. Section \ref{s:conclusion} presents a summary of our findings and discusses possible future endeavours.

{\it At the time of review, it has come to our attention that \cite{Hartnoll:2022rdv} discusses infinite oscillations in a scalar field model and \cite{An:2022lvo} presents results for transitions between different Kasner epochs using a top-down, scalar holographic superconductor model. }


\section{Holographic p-wave superconductor}
\label{s:model}

\subsection{Action and equations of motion}
\label{ss:action_eom}
The model used is a $(3+1)$-dimensional, $SU(2)$ Yang-Mills theory with Einstein-Hilbert and cosmological constant terms allowing us to obtain asymptotically anti-de Sitter (AdS) geometry. The action is
\begin{equation}
\label{eqn:model_Action}
I = \frac{1}{\kappa_{(4)}^{2}}\int d^{3+1}x \sqrt{- g} \left[ R -2 \Lambda - \frac{1}{4} \text{Tr}\left[F_{\mu \nu} F^{\mu \nu} \right]\right] = \int d^{3+1}x \sqrt{-g} \mathcal{L}\,,
\end{equation}
with Lagrangian $\mathcal{L}$ and field strength
\begin{equation}
\label{eqn:model_field_strength}
F^{a}_{\mu \nu} = \nabla_{\mu} A^{a}_{\nu} -  \nabla_{\nu} A^{a}_{\mu} + g_{\text{YM}} \epsilon^{abc}A^{b}_{\mu} A^{c}_{\nu} \,.
\end{equation}
Here, $R$ is the Ricci scalar, $\Lambda = -3/L^{2}$ is the cosmological constant with $L$ the radius of curvature of AdS, $g$ is the determinant of the metric, $\kappa_{(4)}$ is the four-dimensional gravitational constant, $g_{\text{YM}} = \hat{g}_{\text{YM}}/\kappa_{(4)}$ where $\hat{g}_{\text{YM}}$ is the standard Yang-Mills coupling and $\epsilon^{abc}$ is the Levi-Civita symbol. $A_{\mu}^{a}$ represent the Lie-algebra valued gauge fields, defined in form notation as $A = A^{a}_{\mu}\tau^{a} dx^{\mu}$ where $\tau^{a}$ are the generators of the $SU(2)$ algebra defined by the Pauli matrices, $\sigma^{a}$, as $\tau^{a} = \sigma^{a}/2i$. In the above, ``$\text{Tr}$" refers to the trace over Lie indices and in the convention used, $\Tr[\tau^{a} \tau^{b}] = \frac{1}{2}\delta^{ab}$. For example $\frac{1}{4}\text{Tr}\left[ F_{\mu \nu} F^{\mu \nu} \right] = \frac{1}{8} \sum_{a} F^{a}_{\mu \nu} F^{a \mu \nu}$, where for $SU(2)$, the three generators are denoted by $a = 1,2,3$ and satisfy Lie bracket $[\tau^{a}, \tau^{b}] = \epsilon^{abc} \tau^{c}$. Under the identification of $\hat{A}_{\mu} = A_{\mu}/g_{\text{YM}}$ we may think of $g_{\text{YM}}$ as a measure of the backreaction: for large $g_{\text{YM}}$ we enter the probe limit. This limit essentially scales away the effect of the gauge field such that it has negligible contributions to the gravitational equations of motion.

Varying the action of \eqref{eqn:model_Action} with respect to the metric, $g_{\mu \nu}$, and the Yang-Mills gauge field, $A_{\mu}$, the resulting equations of motion are
\begin{equation}
\label{eqn:model_einstein_eom}
R_{\mu \nu} - \frac{1}{2}g_{\mu \nu} R + \Lambda g_{\mu \nu} = T_{\mu \nu}\,,
\end{equation}
\begin{equation}
\label{eqn:model_matter_eom}
D_{\mu} F^{\mu \nu} = 0\,,
\end{equation}
where
\begin{equation}
\label{eqnn:model_EMT_alt}
T_{\mu \nu} =  \frac{1}{2}\text{Tr}\left[F_{\mu \gamma} {F_{\nu}}^{\gamma} \right] - \frac{1}{8}g_{\mu \nu} \text{Tr}\left[F_{\gamma \rho}F^{\gamma \rho}\right]\,.
\end{equation}
Here we define the gauge covariant derivative as
\begin{equation}
\label{eqn:model_cov_derivative}
D_{\mu} = \nabla_{\mu} + g_{\text{YM}}[ A_{\mu}, \cdot] \,.
\end{equation}
In explicit Lie index form, equation \eqref{eqn:model_matter_eom} is 
\begin{equation}
\label{eqn:model_matter_eom_lie_indices}
D_{\mu} F^{a \mu \nu} = \nabla_{\mu} F^{a \mu \nu} + g_{\text{YM}} \epsilon^{abc} A_{\mu}^{b} F^{c \mu \nu} = 0.
\end{equation}

In order to solve the equations of motion, we adopt the following radial direction (labelled coordinate $z$) field ans\"atze for the gauge field
\begin{equation}
\label{eqn:model_gauge_field_ansatz}
A = A^{a}_{\mu}\tau^{a} dx^{\mu} = A^{3}_{t} \tau^{3} dt + A^{1}_{x} \tau^{1} dx = \phi(z) \tau^{3} dt + \omega(z) \tau^{1} dx \,,
\end{equation}
and also the metric
\begin{equation}
\label{eqn:model_metric_ansantz}
ds^2 =  \frac{1}{z^2}\left(-f(z) e^{-\chi(z)} dt^2 + \frac{1}{f(z)} dz^2 + h(z)^2 dx^2 + \frac{1}{h(z)^2} dy^2  \right)\,.
\end{equation}

Inserting these ans\"atze into equations \eqref{eqn:model_einstein_eom} and \eqref{eqn:model_matter_eom} produces five individual equations of motion for the fields
\begin{align}
\label{eqn:model_all_eom_using_ansatz}
\phi '' &= \frac{\omega ^2 \phi  g_{\text{YM}}^2}{f h^2}-\frac{\chi ' \phi '}{2}
\\
\omega'' &= -\frac{e^{\chi } \omega  \phi ^2 g_{\text{YM}}^2}{f^2}-\frac{f' \omega '}{f}+\frac{2 h' \omega '}{h}+\frac{\chi ' \omega '}{2}
\\
\begin{split}
h'' &= \frac{e^{\chi } \omega ^2 z^2 \phi ^2 g_{\text{YM}}^2}{8 f^2 h}-\frac{f' h'}{f}+\frac{h' \chi '}{2}+\frac{2 h'}{z}+\frac{\left(h'\right)^2}{h}-\frac{z^2 \left(\omega '\right)^2}{8 h}
\end{split}
\\
\begin{split}
f' &= \frac{e^{\chi } \omega ^2 z^3 \phi ^2 g_{\text{YM}}^2}{8 f h^2}+\frac{f z^3 \left(\omega '\right)^2}{8 h^2}+\frac{f z \left(h'\right)^2}{h^2}+\frac{3 f}{z}-\frac{3}{L^2 z}+\frac{1}{8} e^{\chi } z^3 \left(\phi '\right)^2
\end{split} 
\\
\chi' &= \frac{e^{\chi } \omega ^2 z^3 \phi ^2 g_{\text{YM}}^2}{8 f^2 h^2}+\frac{f'}{f}+\frac{3}{f L^2 z}-\frac{e^{\chi } z^3 \left(\phi '\right)^2}{8 f}+\frac{z^3 \left(\omega '\right)^2}{8 h^2}+\frac{z \left(h'\right)^2}{h^2}-\frac{3}{z}
\end{align}
A non-trivial $\omega(z)$ profile is responsible for introducing the condensate, $\langle J^{x}_{1} \rangle$, to the model since it breaks the $U(1)$ subgroup symmetry associated to rotations around $\tau^{3}$. In other words, a non-zero $\omega(z)$ picks out the $x$ direction as special and breaks the rotational symmetry in the $x-y$ plane. Naturally, the metric function $h(z)$ accounts for this symmetry breaking in the dual description. The chemical potential is associated with the $U(1)$ symmetry generated by $\tau^{3}$. $\phi(z)$ can be thought as the field dual to the chemical potential, appearing as the field charged under this $U(1)$ symmetry \cite{Gubser:2008wv,Ammon:2009xh,Arias:2012py}.

\subsection{Boundary conditions}

The equations of motion using this ansatz enjoy the following scaling symmetries
\begin{subequations}
\begin{equation}
\label{eqn:model_scaling_sym1}
L \to \alpha_{1} L \,, \quad f \to \frac{1}{\alpha_{1}^2}f \,, \quad g_{\text{YM}} \to \frac{1}{\alpha_{1}} g_{\text{YM}}\,, \quad e^{\chi} \to \frac{1}{\alpha_{1}^2} e^{\chi} \,.
\end{equation}
\begin{equation}
\label{eqn:model_scaling_sym3}
z \to \alpha_{3} z \,, \quad \omega \to \frac{1}{\alpha_{3}} \omega \,, \quad \phi \to \frac{1}{\alpha_{3}} \phi  \,.
\end{equation}
\begin{equation}
\label{eqn:model_scaling_sym2}
e^{\chi} \to \alpha_{2}^2 e^{\chi} \,, \quad \phi \to \frac{1}{\alpha_{2}} \phi \,.
\end{equation}
\begin{equation}
\label{eqn:model_scaling_sym4}
\omega \to \alpha_{4} \omega, \quad h \to \alpha_{4} h  \,.
\end{equation}
\end{subequations}

The symmetries \eqref{eqn:model_scaling_sym1} and \eqref{eqn:model_scaling_sym3} allow us to take $z_{h} = L = 1$. The others also allow us to scale the $\chi$ and $h$ fields such that they take their necessary boundary values: $\chi(z=0) = 0$ and $h(z=0) = 1$. At the boundary we return to AdS spacetime (i.e. we have an asymptotically AdS bulk spacetime) which defines these conditions.

The procedure of numerically obtaining the field solutions from the equations of motion begins by producing series expansions of the fields at the horizon, $z = z_{h} = 1$, and at the UV boundary, $z = 0$. To be precise we only integrate up to a small cut-off value of $z$ for the UV solutions, denoted as $\epsilon$. Analogously we integrate to $z = 1-\delta$ at the horizon, for small $\delta$.
The horizon series takes the following form
\begin{align}
\label{eqn:model_horizon_series}
f &= f_{h1} (z - z_{h}) + f_{h2} (z - z_{h})^2 + \dots \nonumber\\
\chi &= \chi_{h0} + \chi_{h1}(z - z_{h}) + \chi_{h2} (z - z_{h})^2 + \dots \nonumber\\
h &= h_{h0} + h_{h2} (z - z_{h})^2 + \dots\\
\omega &= \omega_{h0} + \omega_{h2}(z - z_{h})^2 + \omega_{h3} (z - z_{h})^3 + \dots \nonumber\\
\phi &= \phi_{h1} (z - z_{h}) + \phi_{h2} (z - z_{h})^2 + \dots  \nonumber
\end{align}
Here $f$ vanishes at $z=z_{h}$ by definition of the black hole event horizon, as does $\phi$ to ensure we have a finite norm of the gauge field strength squared. Substituting these series solutions into the equations of motion and solving order by order, the field functions are completely determined by four parameters at the horizon: $\phi_{h1}$, $\omega_{h0}$, $h_{h0}$, $\chi_{h0}$. Using symmetry \eqref{eqn:model_scaling_sym2}, we choose $\chi_{h0} = 1$ throughout and rescale the necessary quantities when required, to ensure $\chi(z=0) = 0$ i.e. we set $\alpha_{2} = e^{-\chi_{b0}/2}$ with $\chi_{b0}$ defined below in \eqref{eqn:model_chi_uv_expansion}. Repeating the same idea for the UV boundary expansion around $z = 0$, we find
\begin{subequations}
\begin{align}
f &= 1 + f_{b3} z^3 + \mathcal{O}(z^4) \label{eqn:model_f_uv_expansion} \\
\chi &= \chi_{b0} + \mathcal{O}(z^4)  \label{eqn:model_chi_uv_expansion} \\
h &= h_{b0} + h_{b3} z^3 + \mathcal{O}(z^4)   \label{eqn:model_h_uv_expansion} \\
\omega &= \omega_{b0} + \omega_{b1} z  + \mathcal{O}(z^2)  \label{eqn:model_omega_uv_expansion} \\
\phi &= \phi_{b0} + \phi_{b1} z +  \mathcal{O}(z^2) \label{eqn:model_phi_uv_expansion}
\end{align}
\end{subequations}
All higher order terms in $z$ have coefficients that are constructed from the eight UV coefficient parameters listed in equations \eqref{eqn:model_f_uv_expansion}-\eqref{eqn:model_phi_uv_expansion}. 

With the $\chi$ scaling symmetry allowing us to set $\chi_{h0} = 1$, we now look to set the horizon parameters $\phi_{h1}$ and $h_{h0}$ such that we are left with a one-dimensional parameter space of solutions to explore, those being controlled by $\omega_{h0}$. This requires two additional conditions. First,  we require a vanishing source of the $\omega$ field, and in the chosen quantisation provided by the AdS/CFT correspondence, this corresponds to $\omega_{b0} = 0$ from the UV expansion. The correspondence also implies that the expectation value of the dual operator is identified as $\langle J^{x}_{1} \rangle = \omega_{b1}$. This is our condensate. Secondly,  since we return to AdS spacetime at the boundary cut-off, we require that the anisotropy function $h(z)$ become unity there\footnote{In our numerical practice, we make use of the $\chi$ symmetry equation but not the $h$ symmetry equation, instead choosing to ``shoot" for the $h$ function's boundary value. }. These two conditions serve as shooting parameters and root finding algorithms in Mathematica \cite{Mathematica_ref} for example, readily produce solutions. Additionally, the AdS/CFT dictionary states that the chemical potential, $\mu$, and charge density, $\rho$, are identified with coefficients of the $\phi$ UV expansion such that $\mu = \phi_{b0}$, $\rho = \phi_{b1}$.

To establish where the condensate becomes non-trivial, the temperature must be defined. This is the Hawking temperature of the black hole described by equation \eqref{eqn:model_metric_ansantz} and can be obtained through periodicity arguments of the metric's Euclidean signature 
\begin{equation}
\label{eqn:model_temperature}
T = \left. \frac{|f'(z)| e^{-\chi(z)/2}}{4\pi} \right|_{z=z_{h}} \,.
\end{equation}
Using the horizon expansion, this can be explicitly written as\footnote{This can be also be achieved directly using $T = \kappa/2\pi$ where $\kappa$ is the surface gravity defined as $\kappa^2 = - \frac{1}{2}\nabla^{\mu} \xi^{\nu}\nabla_{\mu} \xi_{\nu}$ and $\xi$ is the Killing vector. }
\begin{equation}
\label{eqn:model_temperature_with_coefs}
T = \frac{e^{-\frac{\chi_{h0}}{2}} \left(z_{h}^4 e^{\chi_{h0}} \phi_{h1}^2 -\frac{24}{L^2}\right)}{32 \pi z_{h}} \,.
\end{equation}
Overall, the model is simply determined by two parameters: $T/\mu$, the dimensionless temperature, and $g_{\text{YM}}$, the coupling parameter of the matter fields. Having introduced the general method of acquiring solutions based on the boundary conditions, the following sections proceed to analyse the function content of said solutions, starting with the exterior of the black hole.

\section{Thermodynamics}
\label{s:ext_behaviour}

\subsection{Phase transitions}
\label{ss:field_and_phase}

\begin{figure}[h]
    \centering
    \begin{minipage}{0.5\textwidth}
        \centering
        \includegraphics[width=0.9\textwidth]{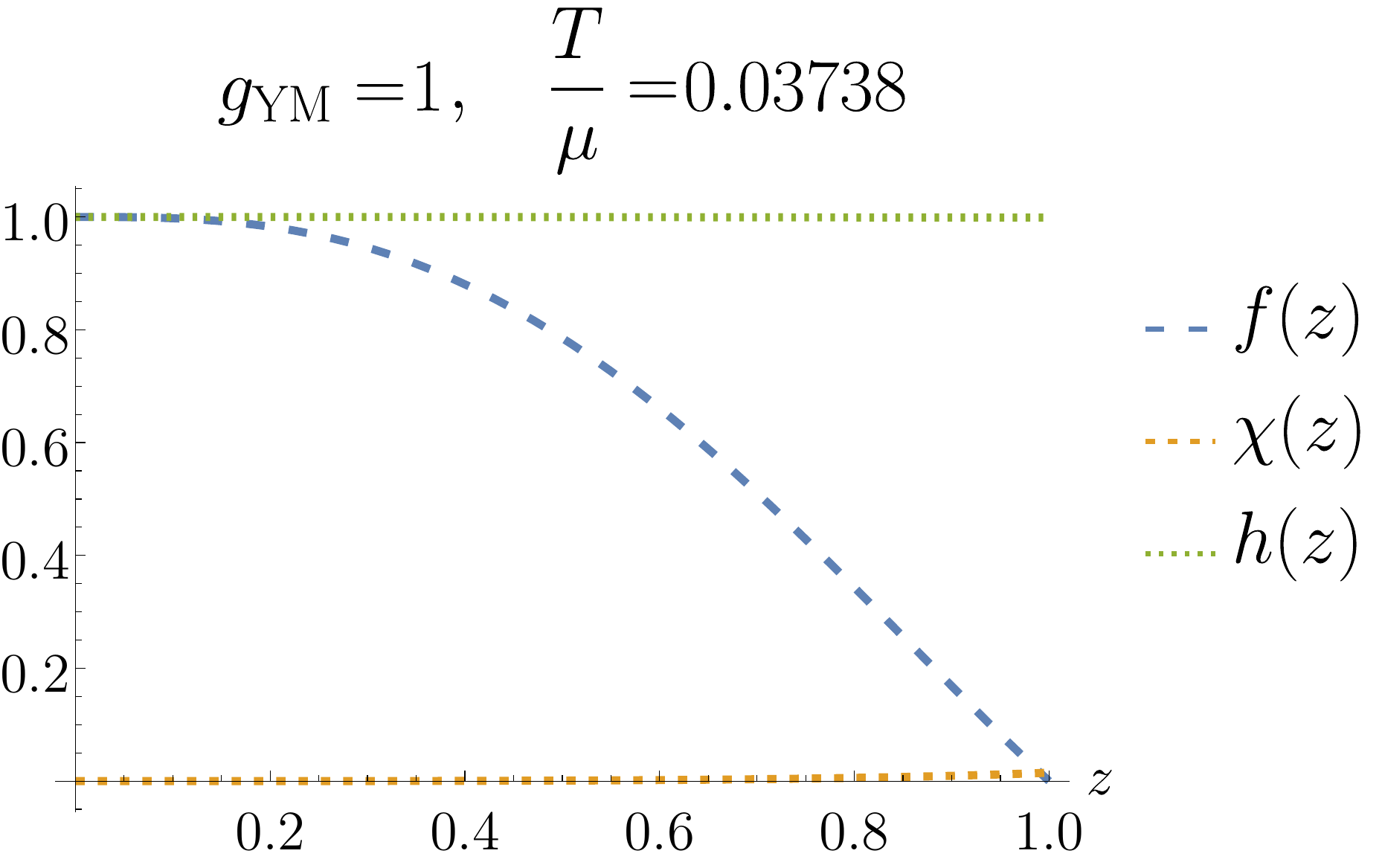} 
    \end{minipage}\hfill
    \begin{minipage}{0.5\textwidth}
        \centering
        \includegraphics[width=0.9\textwidth]{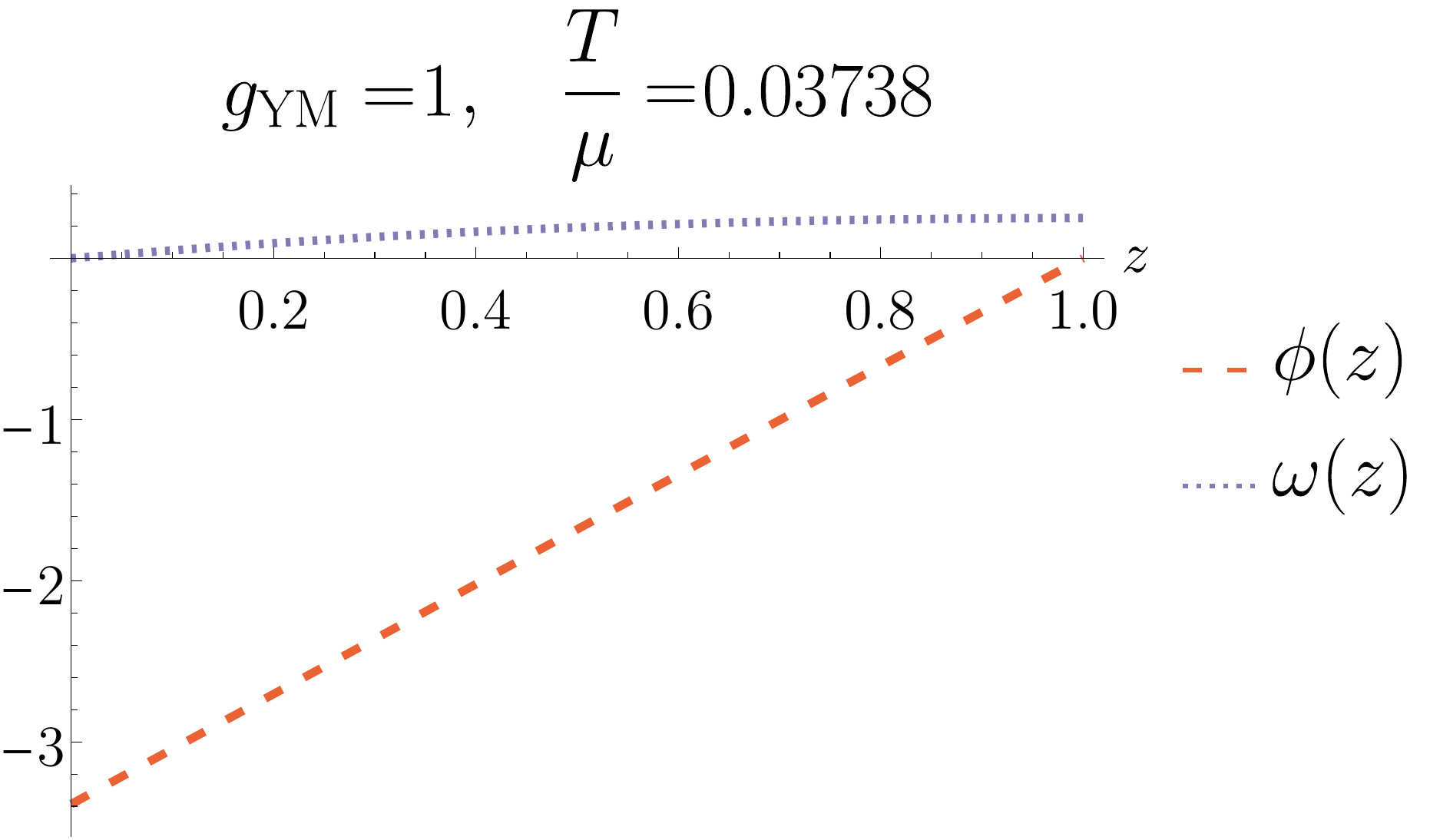}
    \end{minipage}
	\caption{Plots of the matter and metric functions, for $T/\mu = 0.03738$ and $g_{\text{YM}} = 1$. The functions take on a simplified form close to critical temperature  $T_{c}/\mu \approx 0.03748$. $\phi(z)$ becomes approximately linear, $\chi(z) \approx 0$ and $h(z) \approx 1$, while $\omega(z)$ is small and $f(z)$ has polynomial behaviour. Naturally when the temperature is raised above its critical value, the phase changes from superconducting to normal and the Reissner-Nordstr\"{o}m geometry is recovered.}
	\label{fig:met_and_matter_funcs_plot2}
\end{figure}

\begin{figure}[h]
    \centering
    \begin{minipage}{0.5\textwidth}
        \centering
        \includegraphics[width=0.9\textwidth]{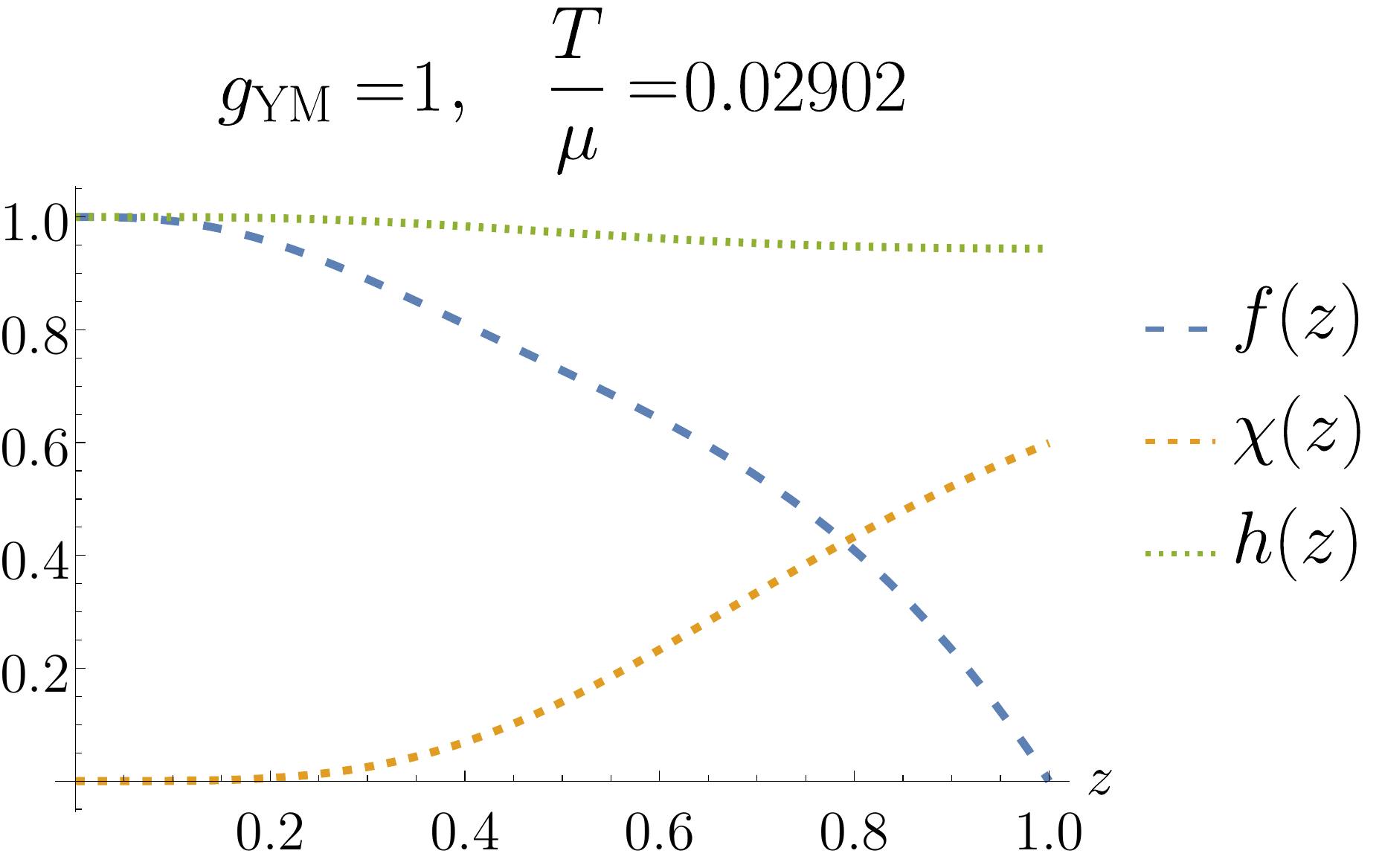} 
    \end{minipage}\hfill
    \begin{minipage}{0.5\textwidth}
        \centering
        \includegraphics[width=0.9\textwidth]{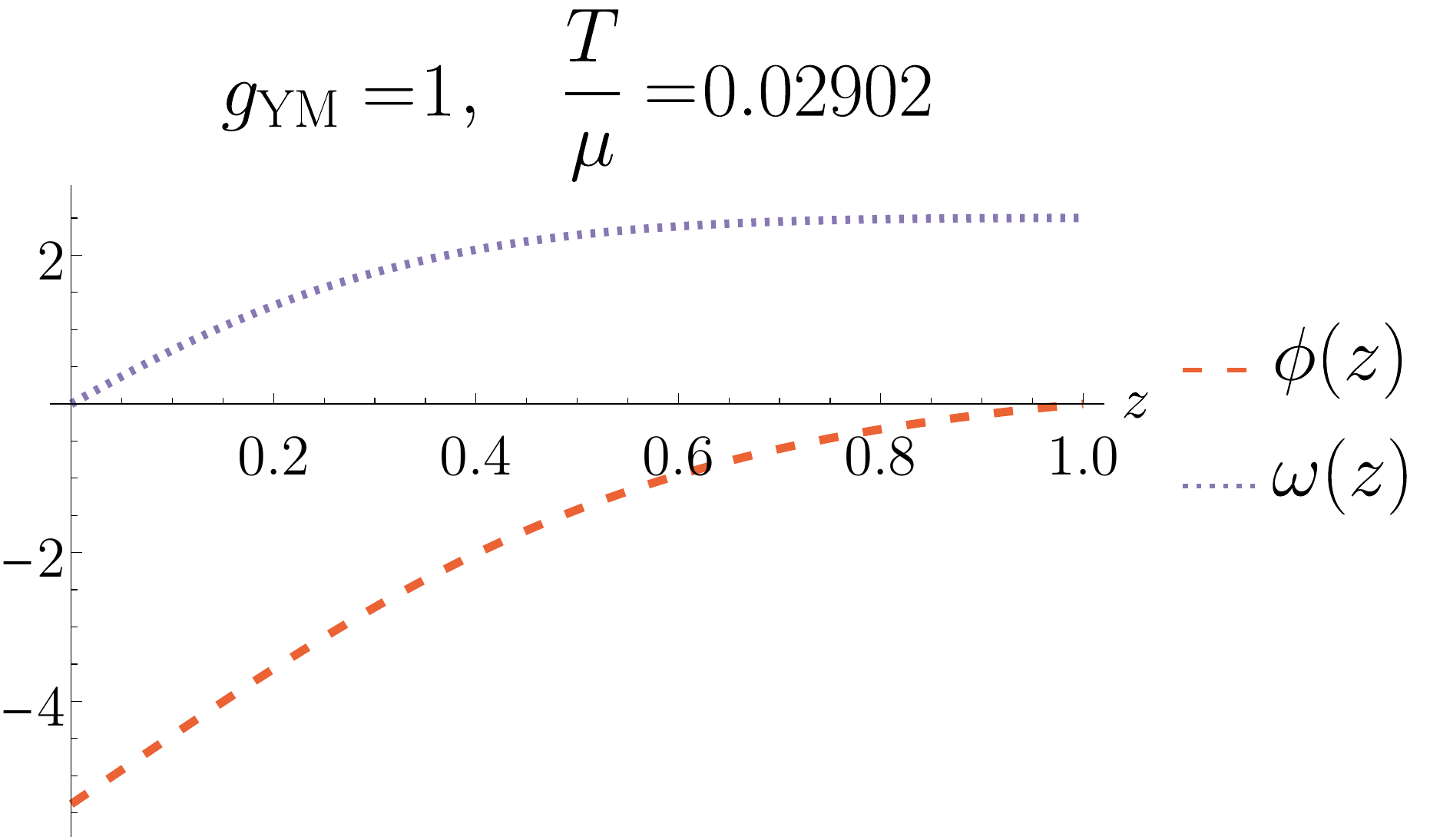}
    \end{minipage}
	\caption{Plots of the matter and metric functions, for $T/\mu = 0.02902$ and $g_{\text{YM}} =1$. The function behaviour far away from critical temperature (and within superconducting phase) is far more complicated. This is especially visible in the non-constant $h(z)$ and $\chi(z)$ and also in the larger value of $\omega(z)$ throughout the exterior. }
	\label{fig:met_and_matter_funcs_plot1}
\end{figure}

Before exploring the interior, the core features of the black hole exterior are studied. We begin by presenting the field behaviour between the UV boundary, $z=0$, and horizon, $z= z_{h} = 1$, at two different temperatures: one close to critical temperature shown in Figure~\ref{fig:met_and_matter_funcs_plot2} and one far away shown in Figure \ref{fig:met_and_matter_funcs_plot1}, both for $g_{\text{YM}} = 1$ (analysis shows that this coupling value permits typical normal-to-superconducting transitions).

At both temperatures chosen, the functions achieve their correct forms at the horizon and UV boundaries: $\phi(z_{h}) = f(z_{h}) = 0$, while $h(0) = 1$, $\chi(0)=0$ and $f(0) = 1$. Analytic temperatures for the $SU(2)$ model exist \cite{Herzog:2009ci} and for these parameter choices the critical temperature is $T_{c}/\mu \approx 0.03748$. It therefore makes sense that Figure \ref{fig:met_and_matter_funcs_plot2} exhibits such a form. Approaching critical temperature is essentially approaching normal phase where the condensate vanishes, $\omega(z) = 0$. When this occurs, we return to a Reissner-Nordstr\"{o}m solution
\begin{gather}
\begin{gathered}
\label{eqn:reiss_nord_sol}
\omega(z) = 0\,, \quad \phi(z) = \mu\left(1 - \frac{ z}{z_{h}}\right) \,, \\
 \chi(z) = 0\,, \quad  h(z) = 1\,, \quad f(z) = \frac{1}{L^2} -  \frac{z^3}{L^2 z_{h}^3} - \frac{\mu^2 z^3}{8 z_{h}} + \frac{\mu^2 z^4}{8 z_{h}^2} \,.
\end{gathered}
\end{gather}

As expected the functions close to critical temperature of Figure \ref{fig:met_and_matter_funcs_plot2} are converging to these solutions. This is clear by the non-linear behaviour of $f(z)$, the constant behaviours of $\chi(z)$ and $h(z)$, the linear behaviour of $\phi(z)$, and the diminishing value of $\omega(z)$ throughout the exterior. As for the lower temperature seen in Figure \ref{fig:met_and_matter_funcs_plot1} the behaviour becomes more complicated since the solution is in superconducting phase. Note that these plots feature the rescaled functions i.e. using symmetry \eqref{eqn:model_scaling_sym2}.

\begin{figure}[htb!]
\begin{center}
\includegraphics[width=11cm, height=6.5cm]{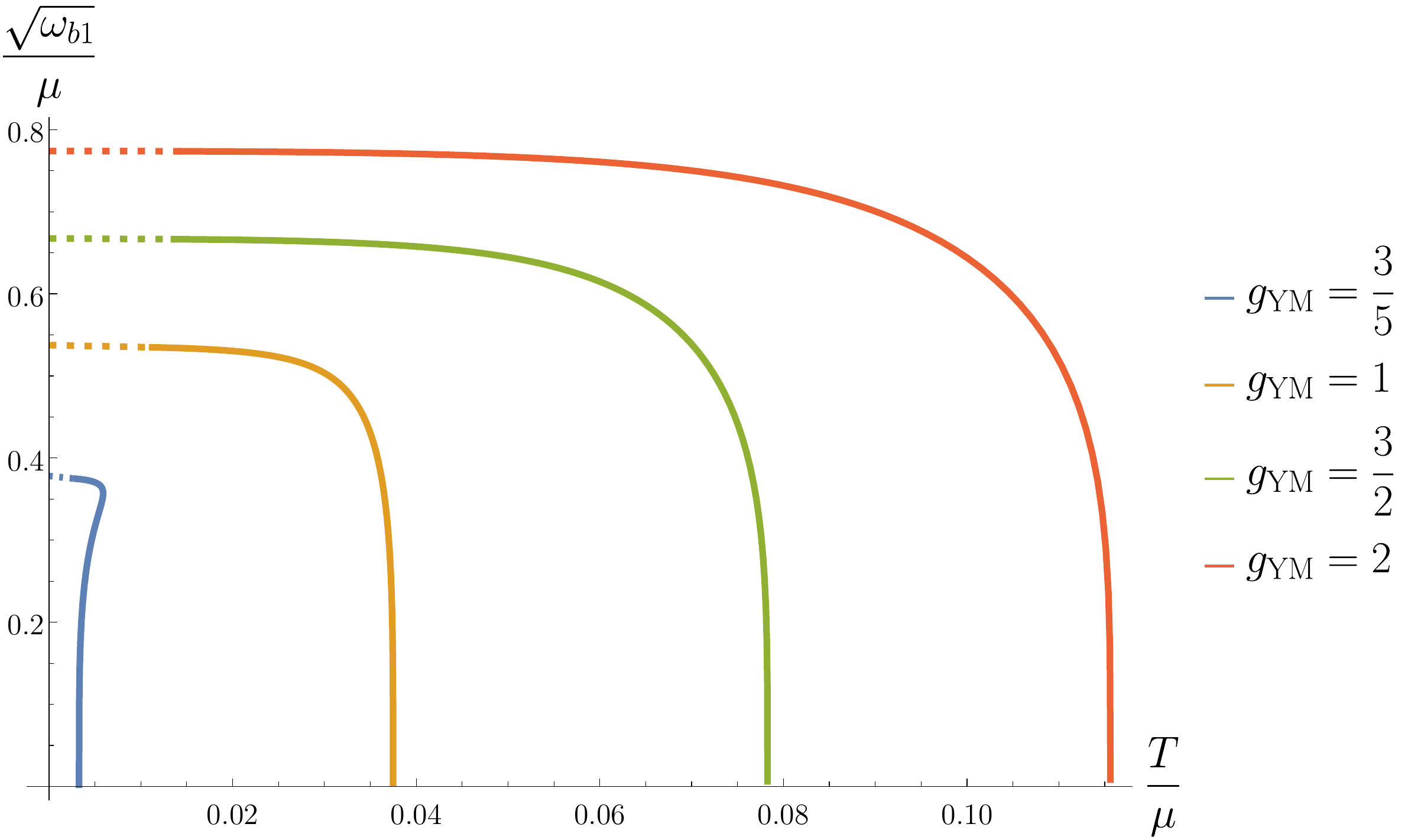}
\end{center}
\caption{Phase diagrams for the four values of $g_{\text{YM}}$ considered. The $g_{\text{YM}} = 1$, $3/2$ and $2$ curves all exhibit similar behaviour, while the the $g_{\text{YM}} = 3/5$ curve becomes multi-valued between certain temperatures. This indicates that a critical value, $g^{c}_{\text{YM}}$, exists and from inspection is $g_{\text{YM}}^{c}  = 0.8075 \pm 0.0025$. Below $g^{c}_{\text{YM}}$ there are first order phase transitions, whilst above $g^{c}_{\text{YM}}$ there will be second order transitions. This can be confirmed by studying the grand potential and entropy. The dashed part of the curves represent extrapolation of the data to reach zero temperature.} 
\label{fig:allphaseplots}
\end{figure}

Having observed the solutions, we now study the phase transitions for four choices of $g_{\text{YM}}$. The main feature found is the existence of a critical $g_{\text{YM}}$ at which the phase transition changes from first to second order. This behaviour has appeared in \cite{Ammon:2009xh,Cai:2013aca} where models of different dimension and Lagrangian\footnote{This model \cite{Cai:2013aca} can actually be identified with the present model under specific selection of fields/parameters, see appendix A of \cite{Cai:2021obq}.} were studied respectively\footnote{Note that first order transitions were not found in \cite{Arias:2012py}, however, this was specifically due to the parameter range the authors chose to study: a range in which their $g_{\text{YM}}$ values were greater than where we found first order transitions.}. Plots of the condensate vs. temperature are made in Figure \ref{fig:allphaseplots} for the four $g_{\text{YM}}$ values: $3/5$, $1$, $3/2$, $2$. Since $g_{\text{YM}} = 3/5$ appears multivalued and $g_{\text{YM}} = 1$ does not, the critical value should obey $  3/5 < g^{c}_{\text{YM}} < 1$ and upon inspection we indeed find $g_{\text{YM}}^{c}  = 0.8075 \pm 0.0025$.

\FloatBarrier
\subsection{Grand potential and entropy analysis}
\label{ss:gp_and_entropy_analysis}

To better understand the type of phase transition we see at these $g_{\text{YM}}$, the grand  potential $\Omega$, is calculated via the Euclidean action. The method outline is as follows. There are two solutions to the model which may be studied. The first is for normal phase where there is zero condensate, $\langle J_{1}^{x} \rangle = 0$. This corresponds to the Reissner-Nordstr\"{o}m solution seen in \eqref{eqn:reiss_nord_sol}. The second is for condensed phase where there is non-zero condensate $\langle J_{1}^{x} \rangle \neq 0$, based on the results of our full numerics. By generating $\Omega$ for both of these solutions, we can see which is physically favourable by identifying which takes the smaller value.  Additionally the black hole entropy, $S$, is also calculated to help visualise the phase transition \cite{Ammon:2009xh}.

The grand potential, $\Omega$, is simply the Euclidean action, $I^{E}$, multiplied by the temperature. In order to produce a well defined Dirichlet boundary problem for calculating the Euclidean action, we must include the standard Gibbons-Hawking boundary term, while to handle the divergences from both the bulk and boundary terms, a counter term must also be added. Hence, the total Euclidean action takes the form
\begin{equation}
\label{eqn:eb_total_euclidean_action}
I^{\text{E}}_{\text{total}} = \underbrace{ - \int d^{4} x \sqrt{-g} \mathcal{L}}_{I^{\text{E}}_{\text{bulk}}} \underbrace{- \frac{2}{\kappa_{(4)}^2} \int d^{3}x \sqrt{-\gamma}  \mathcal{K}}_{I^{\text{E}}_{\text{GH}}} \underbrace{- \frac{4}{\kappa_{(4)}^2}  \int d^{3}x \sqrt{-\gamma}}_{I^{\text{E}}_{\text{ct}}} \,,
\end{equation}
where $I^{\text{E}}_{\text{bulk}}, I^{\text{E}}_{\text{GH}}$ and $I^{\text{E}}_{\text{ct}}$ are the Euclidean bulk, Gibbons-Hawking and counter terms respectively. We define $\gamma_{ij}$ as the induced boundary metric on a fixed $z$ hypersurface, $z_{b}$, and the determinant of $\gamma_{ij}$ is denoted $\gamma$. Also, $\mathcal{K}$ is the trace of the extrinsic curvature, defined as $\mathcal{K} = \gamma^{ij} K_{ji}$ where $K_{ij}$ is the extrinsic curvature (second fundamental form). 

Beginning with the bulk term, we may write it as
\begin{equation}
\label{eqn:eb_euclid_s_bulk}
I^{\text{E}}_{\text{bulk}} = -\int d^{4}x \sqrt{- g} \frac{1}{\kappa_{(4)}^{2}}\left[ R -2 \Lambda - \frac{1}{4} \text{Tr}\left[F_{\mu \nu} F^{\mu \nu} \right]\right] = - \beta V \int dz \sqrt{- g} \mathcal{L}\,,
\end{equation}
with the $d^4 x$ now replacing $d^{3+1} x$ having adopted Euclidean signature with time coordinate, $t_{\text{E}}$. Creating the Euclidean action requires that  $t_{\text{E}} \sim t_{\text{E}} + i \beta$ where $\beta$ is the inverse temperature $T = 1/\beta$. In this case, integrating over the Euclidean time direction is integrating over the thermal circle. Since the Lagrangian is independent of $t_{\text{E}}$ then the integration simply produces a factor of $\int dt_{\text{E}} = \beta$. Similarly, the Lagrangian is independent of $x,y$ and so $\int dx dy = V$ is chosen to represent the volume of the two dimensional $(x,y)$-space. 

\begin{figure}[h]
    \centering
    \begin{minipage}[t]{0.49\textwidth}
        \centering
        \includegraphics[width=1\textwidth]{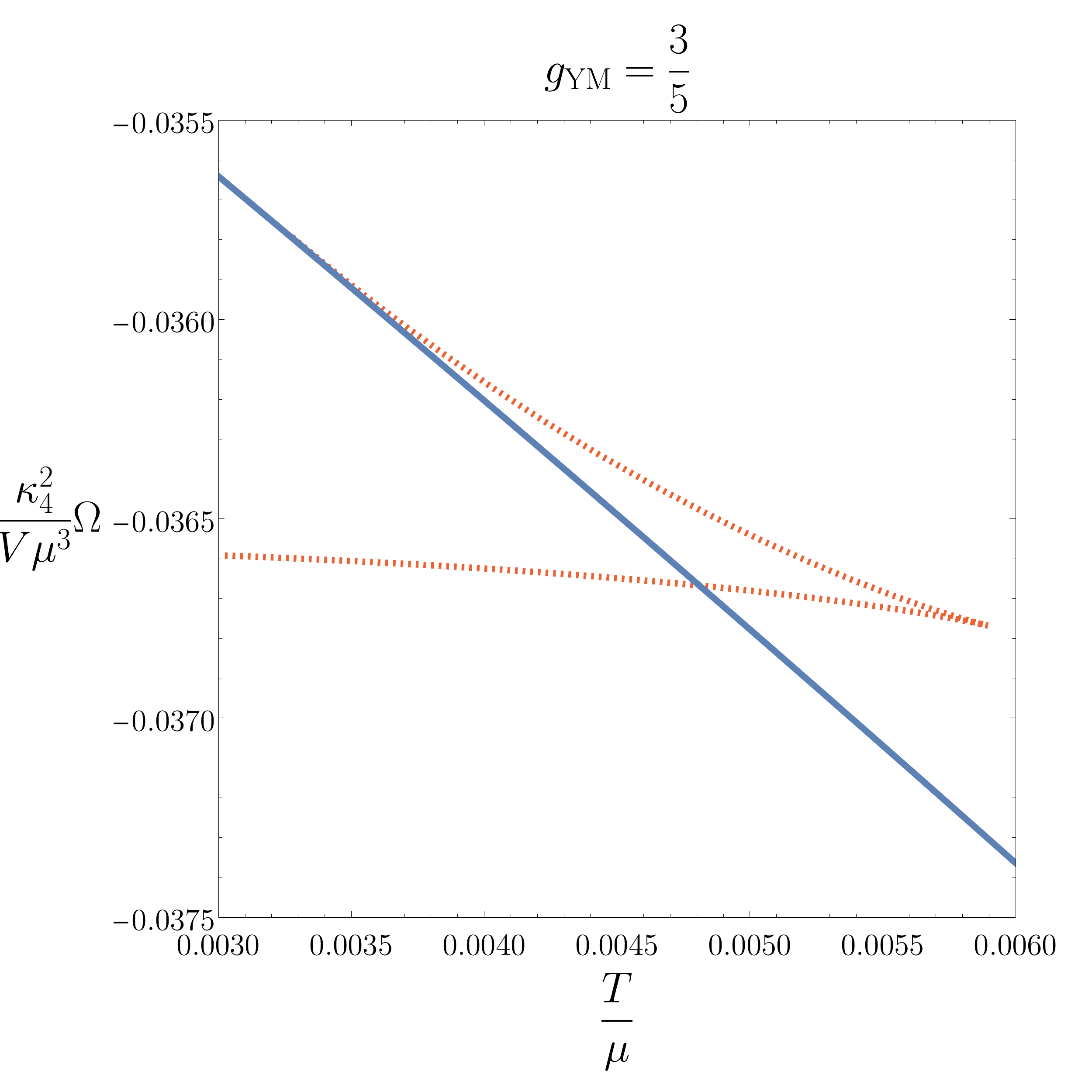} 
        \caption{The grand potential $\Omega$ as a function of temperature for both normal (blue) and condensed (red, dashed) phase with $g_{\text{YM}} = 3/5$. The condensed phase curve exhibits swallow tail behaviour, and the intersection between it and the normal phase curve shows that $\Omega$ is continuous, but non-differentiable. This is characteristic of first order phase transition.}
        \label{fig:gcp1plot}
    \end{minipage}
    \hfill
    \begin{minipage}[t]{0.49\textwidth}
        \centering
        \includegraphics[width=1\textwidth]{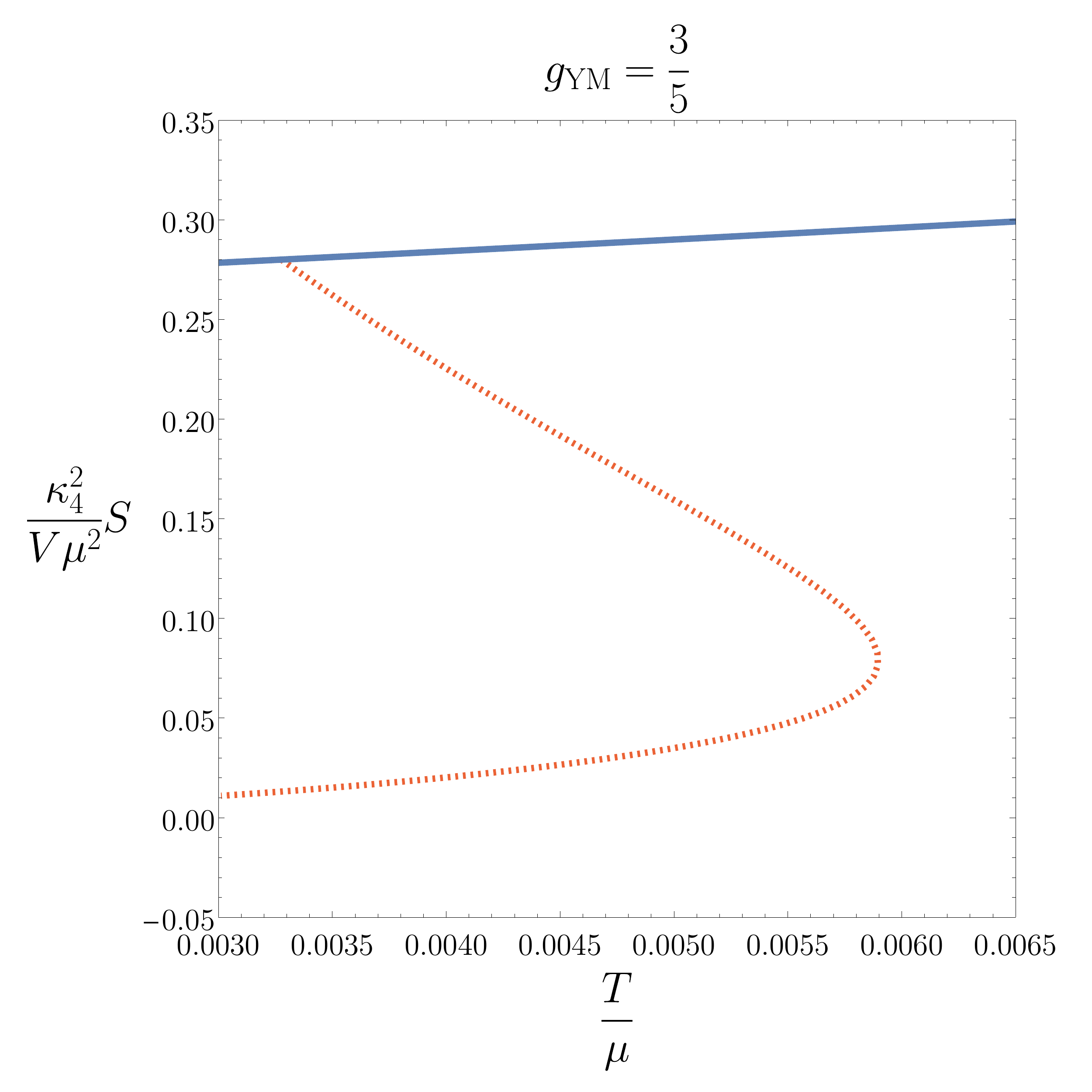}
        \caption{The entropy $S$ as a function of temperature for both normal (blue) and condensed (red, dashed) phase with $g_{\text{YM}} = 3/5$. At $T=T_{c}$, the condensed curve is not continuous as following the normal curve down past $T=T_{c}$, there is discontinuous jump to the lower condensed curve branch. The behaviour exhibited in these plots is therefore that of first order phase transition.}
        \label{fig:ent1plot}
    \end{minipage}
\end{figure}

\begin{figure}[h]
    \centering
    \begin{minipage}[t]{0.49\textwidth}
        \centering
        \includegraphics[width=1\textwidth]{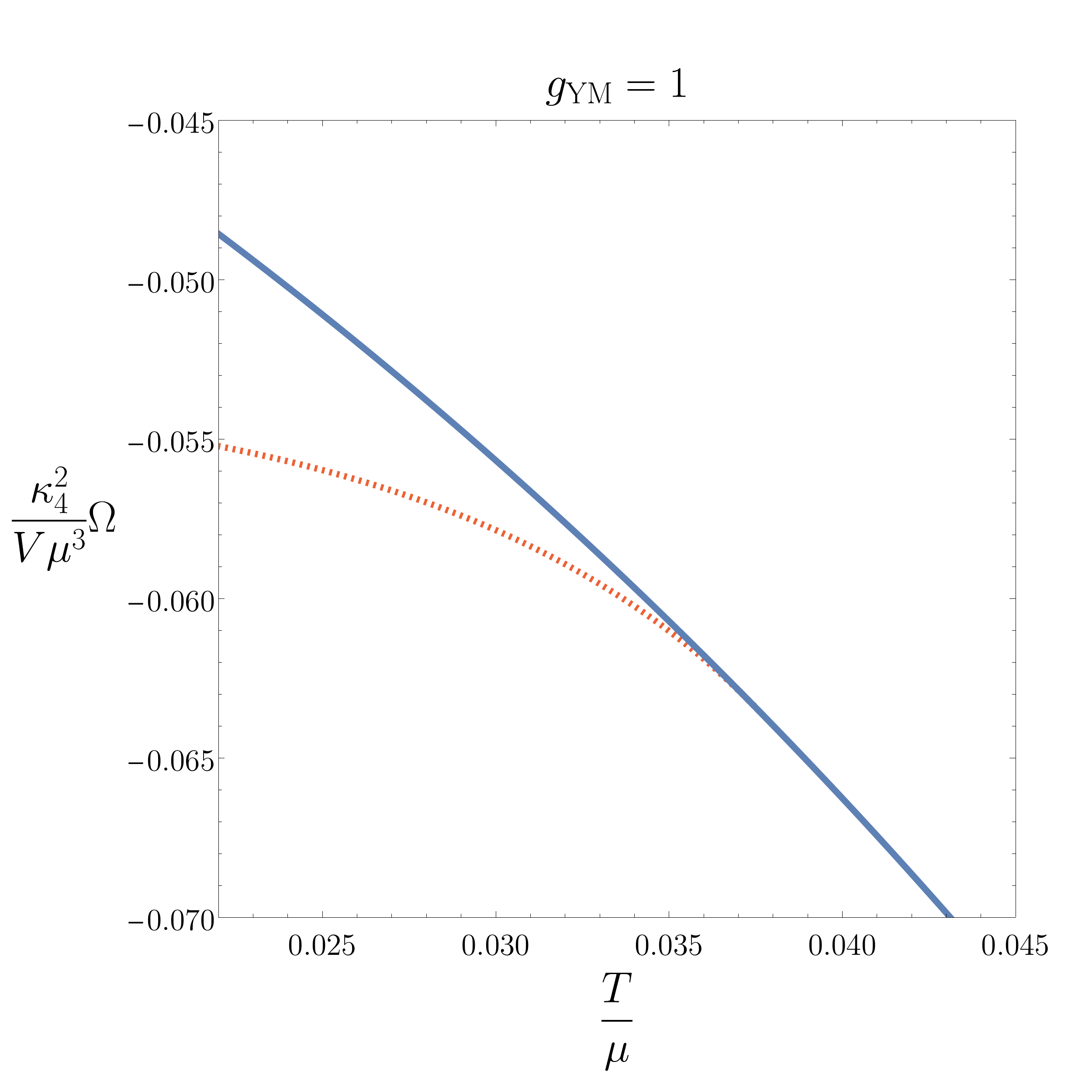} 
        \caption{The grand potential $\Omega$ as a function of temperature for both normal and condensed phase with $g_{\text{YM}} = 1$. The red dashed curve (condensed phase, $\langle J^{x}_{1} \rangle \neq 0$) and the blue curve (normal phase,  $\langle J^{x}_{1} \rangle = 0$) smoothly intersect showing that $\Omega$ is both continuous and differentiable.}
        \label{fig:gcp2plot}
    \end{minipage}
    \hfill
    \begin{minipage}[t]{0.49\textwidth}
        \centering
        \includegraphics[width=1\textwidth]{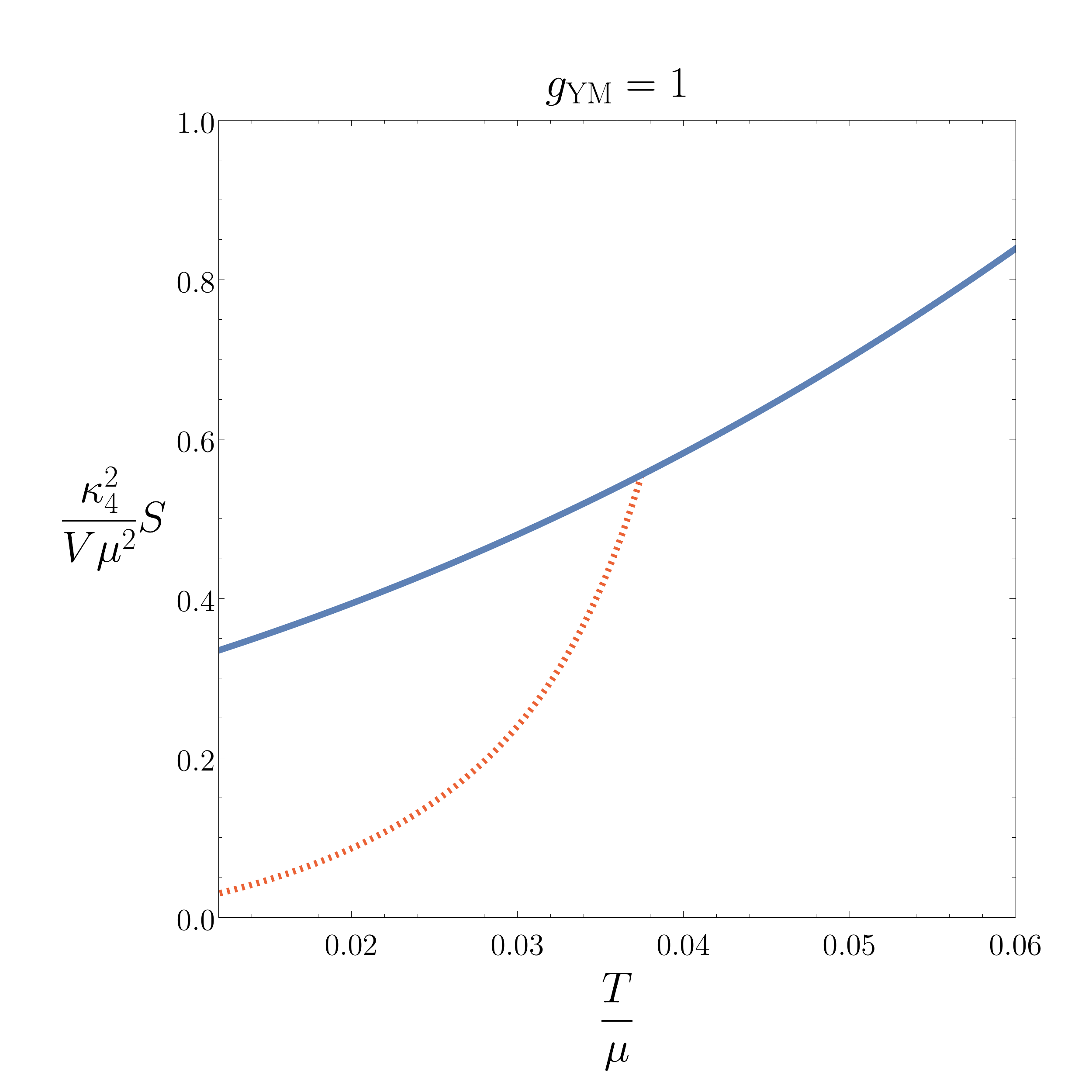}
        \caption{ The entropy $S$ as a function of temperature for both normal (blue) and condensed (red, dashed) phase with $g_{\text{YM}} = 1$. At $T=T_{c}$, the red curve is continuous but non-differentiable. The behaviour exhibited is therefore that of second order phase transition.}
        \label{fig:ent2plot}
    \end{minipage}
\end{figure}

\begin{figure}[h]
    \centering
    \begin{minipage}[t]{0.49\textwidth}
        \centering
        \includegraphics[width=1\textwidth]{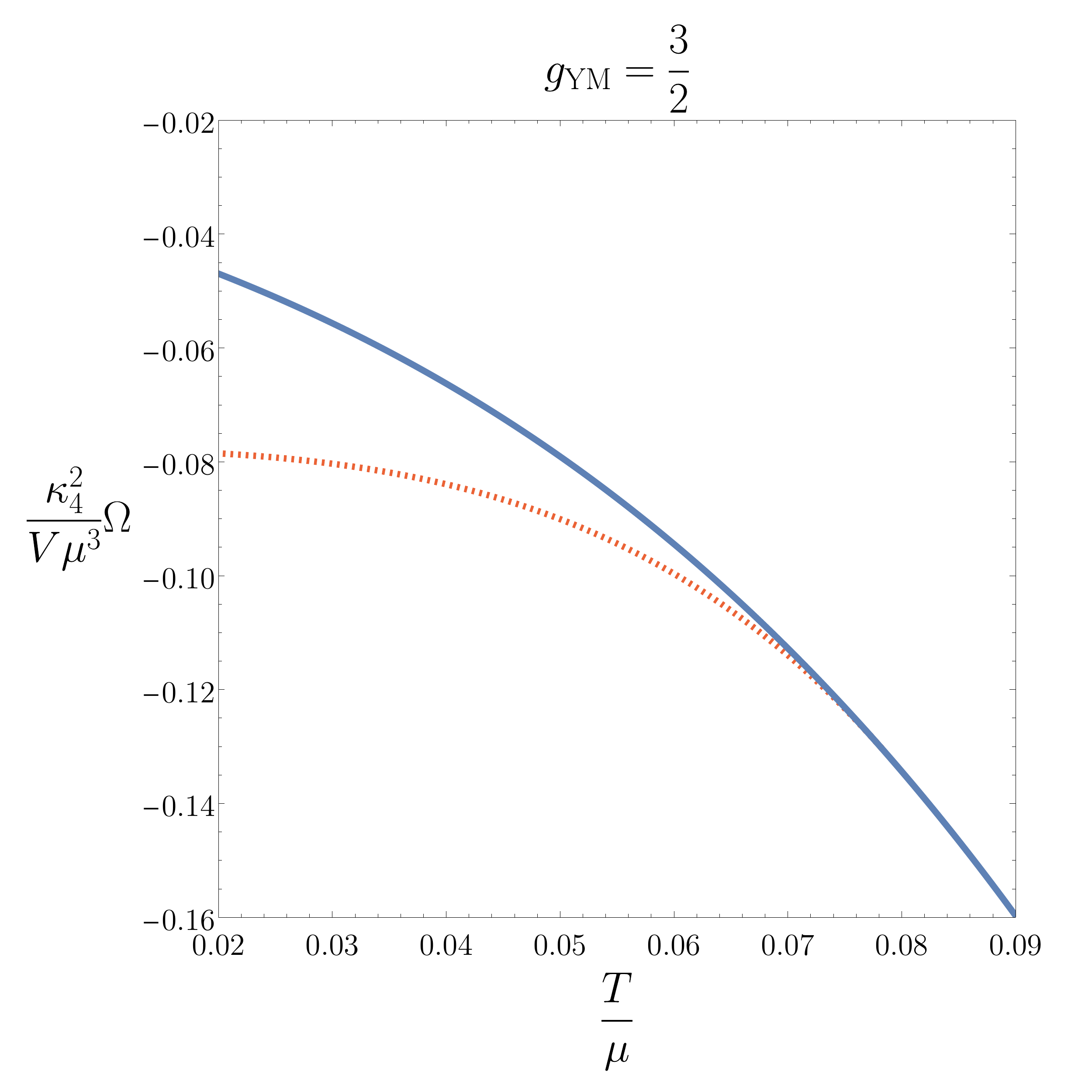} 
         \caption{The grand potential $\Omega$ as a function of temperature for both normal and condensed phase with $g_{\text{YM}} = 3/2$. The red dashed curve (condensed phase, $\langle J^{x}_{1} \rangle \neq 0$) and the blue curve (normal phase,  $\langle J^{x}_{1} \rangle = 0$)  smoothly intersect showing that $\Omega$ is both continuous and differentiable.}
         \label{fig:gcp3plot}
    \end{minipage}
    \hfill
    \begin{minipage}[t]{0.49\textwidth}
        \centering
        \includegraphics[width=1\textwidth]{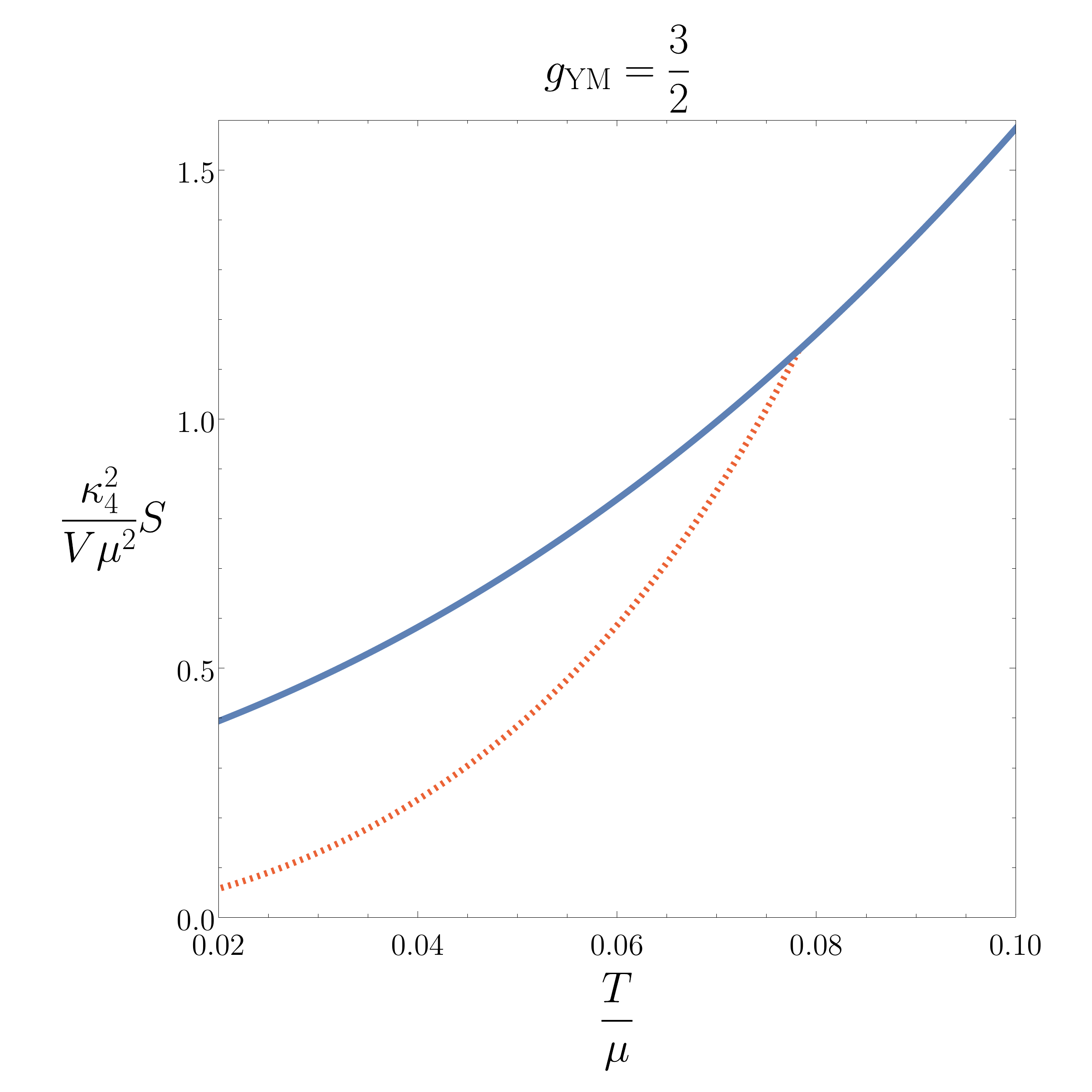}
        \caption{ The entropy $S$ as a function of temperature for both normal (blue) and condensed (red, dashed) phase with $g_{\text{YM}} = 3/2$. At $T=T_{c}$, the red curve is continuous but non-differentiable. The behaviour exhibited is therefore that of second order phase transition.}
        \label{fig:ent3plot}
    \end{minipage}
\end{figure}

\begin{figure}[h]
    \centering
    \begin{minipage}[t]{0.49\textwidth}
        \centering
        \includegraphics[width=1\textwidth]{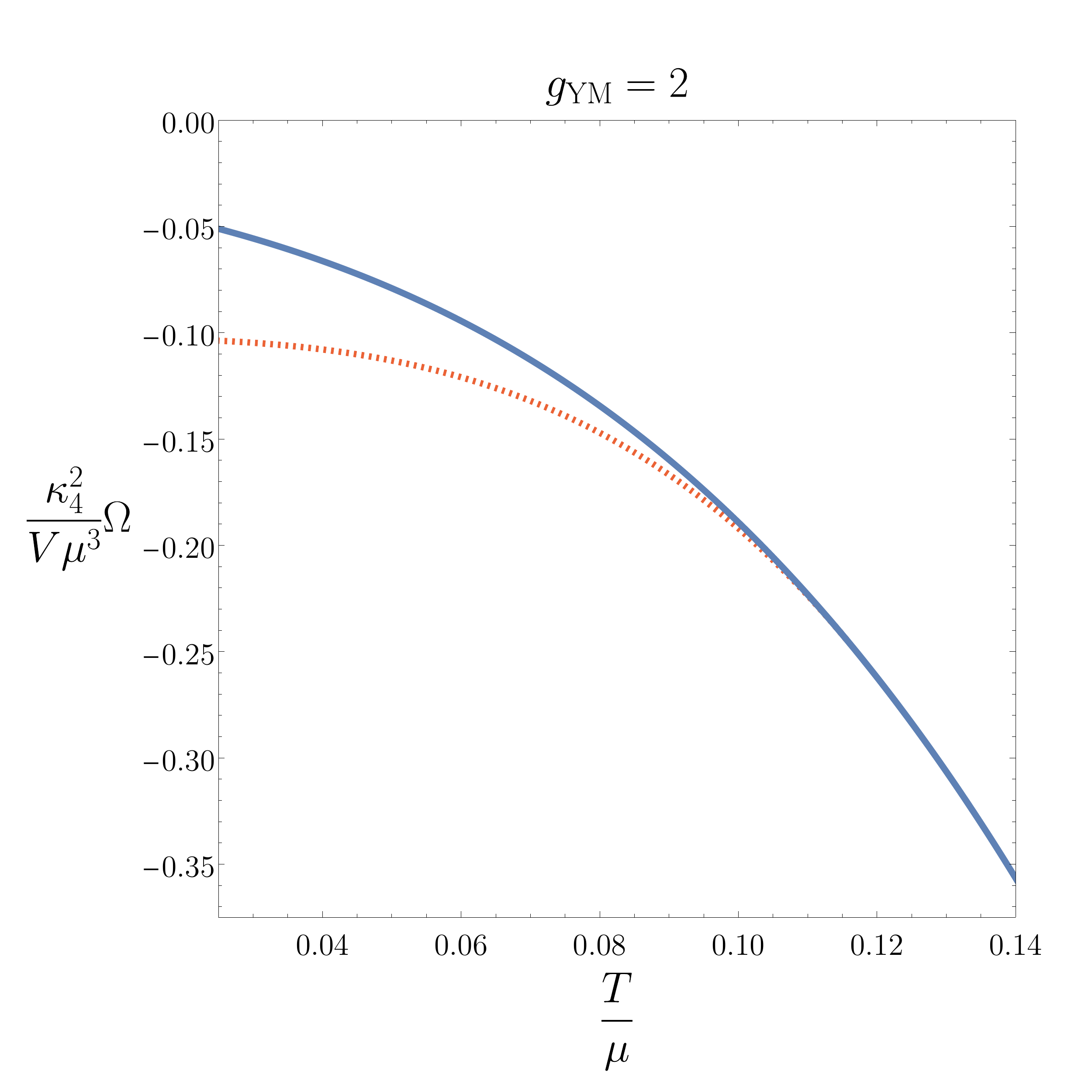} 
         \caption{The grand potential $\Omega$ as a function of temperature for both normal and condensed phase with $g_{\text{YM}} = 2$. The red dashed curve (condensed phase, $\langle J^{x}_{1} \rangle \neq 0$) and the blue curve (normal phase,  $\langle J^{x}_{1} \rangle = 0$) smoothly intersect showing that $\Omega$ is both continuous and differentiable.}
         \label{fig:gcp4plot}
    \end{minipage}
    \hfill
    \begin{minipage}[t]{0.49\textwidth}
        \centering
        \includegraphics[width=1\textwidth]{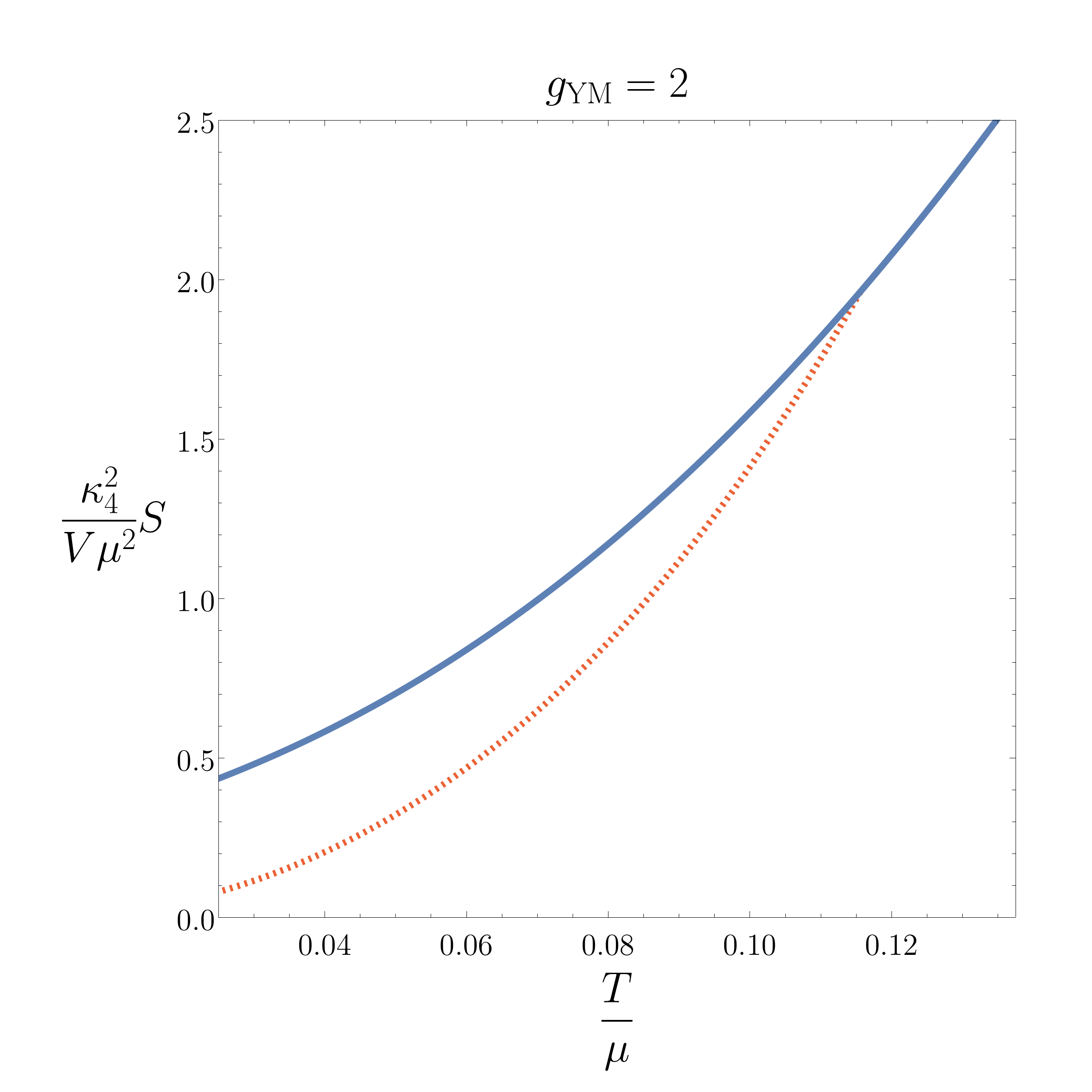}
	    \caption{ The entropy $S$ as a function of temperature for both normal (blue) and condensed phase (red, dashed) with $g_{\text{YM}} = 2$. At $T=T_{c}$, the red curve is continuous but non-differentiable. The behaviour exhibited is therefore that of second order phase transition.}
    	\label{fig:ent4plot}
    \end{minipage}
\end{figure}

To evaluate the integral, it is useful to note that the $yy$ component of the Einstein tensor, $G_{\mu \nu} = R_{\mu \nu} - \frac{1}{2} g_{\mu \nu} R$, has a simple relationship with the Lagrangian $\mathcal{L}$ \cite{Arias:2012py}
\begin{equation}
\label{eqn:eb_yy_einstein_tensor_and_lag}
{G^{y}}_{y} = \frac{1}{2} \left(\kappa_{(4)}^2 \mathcal{L} - R\right)\,.
\end{equation} 
This connection allows us to describe $\mathcal{L}$ purely in terms of metric functions\footnote{Alternatively, one may substitute the equations of motion directly into the Lagrangian to remove second derivatives and achieve the same form as equation \eqref{eqn:eb_lag_with_ansatz_funcs}.}
\begin{equation}
\label{eqn:eb_lag_einstein_relation}
 \mathcal{L} = \frac{2}{\kappa_{(4)}^2}\left[ {G^{\mu}}_{\mu} - \left(g^{tt}G_{tt} + g^{zz} G_{zz} + g^{xx} G_{xx} \right) \right] + \frac{R}{\kappa_{(4)}^2} \,.
\end{equation}
where ${G^{\mu}}_{\mu} = g^{\mu \nu} G_{\nu \mu}$. Using our ansatz of the fields, one can then cast the Lagrangian above as
\begin{equation}
\label{eqn:eb_lag_with_ansatz_funcs}
\mathcal{L} = \frac{z^4}{\kappa_{(4)}^{2} e^{-\chi/2}} \left[ \frac{2 f e^{-\chi/2} }{z^3 h} \left( z h \right)^{'} \right]^{'} \,.
\end{equation}
where $\alpha'$ denotes $d \alpha/ dz$. The integral of \eqref{eqn:eb_euclid_s_bulk} simplifies since the pre-factor of the total derivative in $\mathcal{L}$ is simply the determinant of the metric. Therefore
\begin{align}
\label{eqn:eb_euc_action_evaluated}
I^{\text{E}}_{\text{bulk}} &= -\beta V \int dz \sqrt{- g} \mathcal{L} \\
&= -\frac{\beta V}{\kappa_{(4)}^{2}} \int dz  \frac{ e^{ -\chi/2} }{z^4 }   \frac{z^4}{e^{-\chi/2}} \left[ \frac{2 f e^{-\chi/2} }{z^3 h} \left( z h \right)^{'} \right]^{'}  \\
& = -\frac{\beta V}{\kappa_{(4)}^{2}} \left[ \frac{2 f e^{-\chi/2} }{z^3 h} \left( z h \right)^{'} \right] \bigg|_{z = z_{b}} \label{eqn:eb_euc_action_evaluated2} \,,
\end{align}
where we evaluate the bulk Euclidean action at $z=z_{b}$ which is taken as the boundary of space. After adding the GH term and finally the counterterm, this regulated hypersurface will be removed by taking $z_{b} \to 0$. Moving onto the GH term
\begin{equation}
\label{eqn:eb_gh_term_eval}
I^{\text{E}}_{\text{GH}} = -\frac{2}{\kappa_{(4)}^2} \int d^{3}x \sqrt{-\gamma}  \mathcal{K} = -\frac{\beta V }{ \kappa_{(4)}^2} \frac{e^{-\chi/2}}{z^3}  \left(z f' - f(6 + z \chi') \right)\bigg|_{z = z_{b}} \,,
\end{equation}
and finally the counterterm
\begin{equation}
\label{eqn:eb_ct_term_eval}
I^{\text{E}}_{\text{ct}} = - \frac{4}{\kappa_{(4)}^2}  \int d^{3}x \sqrt{-\gamma} = -\frac{ \beta V}{ \kappa_{(4)}^2}\frac{4 e^{-\chi/2} \sqrt{f}}{z^3} \bigg|_{z = z_{b}} \,.
\end{equation}
Summing equations \eqref{eqn:eb_euc_action_evaluated2}, \eqref{eqn:eb_gh_term_eval} and \eqref{eqn:eb_ct_term_eval} produces the total Euclidean action. Defining $\tilde{I}^{\text{E}}_{\text{total}} = \frac{1}{\beta} I^{\text{E}}_{\text{total}}$ for convenience, we have
\begin{equation}
\label{eqn:eb_gcp_def}
\Omega =  \lim_{z \to z_{b}} \tilde{I}^{\text{E}}_{\text{total}}  = -\frac{V}{\kappa_{(4)}^2} \frac{e^{-\chi/2}}{z^3}\left[\frac{2 f  }{ h} \left( z h \right)^{'} + z f' - f(6 + z \chi') + 4\sqrt{f} \right]\bigg|_{z = z_{b}}\,.
\end{equation}
In order to evaluate $\Omega$, we must take $z_{b}$ as some small cut off. This permits the insertion of the UV boundary expansions into \eqref{eqn:eb_gcp_def}, and afterwards the regulator can be removed by taking $z_{b} \to 0$. Doing so results in the final expression
\begin{equation}
\label{eqn:eb_gcp_exp1}
 \Omega = -\frac{V}{\kappa_{(4)}^2}(f_{b3} + 6 h_{b3})\,.
\end{equation} 
A natural check of this quantity as well as the consistency of conformality on the boundary, comes from analysis of the boundary stress energy tensor \cite{Balasubramanian:1999re,deHaro:2000vlm}. The stress energy tensor for the present model is defined as 
\begin{equation}
\label{eqn:eb_bry_stress_energy_tensor}
\langle T_{ij} \rangle = \lim_{z \to z_{b}} \frac{2}{\sqrt{\gamma}} \frac{\delta \tilde{I}^{\text{E}}_{\text{total}}}{\delta \gamma^{ij}} = -\frac{2  V}{\kappa_{(4)}^2}\lim_{z \to z_{b}}\frac{1}{z}\left(K_{ij} - \mathcal{K} \gamma_{ij} - 2 \gamma_{ij} \right) \,. 	
\end{equation}
Again, using the ans\"atze and substituting in the UV boundary expansion, we obtain the three stress tensor boundary components 

\begin{equation}
\label{eqn:eb_stress_enegy_tt}
\langle T_{tt} \rangle = \frac{ V}{ \kappa_{(4)}^2}(2f_{b3}) \,,
\end{equation}

\begin{equation}
\label{eqn:eb_stress_enegy_xx}
\langle T_{xx} \rangle = \frac{ V}{ \kappa_{(4)}^2}(f_{b3}-6h_{b3}) \,,
\end{equation}

\begin{equation}
\label{eqn:eb_stress_enegy_yy}
\langle T_{yy} \rangle = \frac{ V}{ \kappa_{(4)}^2} (f_{b3} + 6h_{b3} )\,,
\end{equation}
where we have taken the hypersurface $z_{b} \to 0$. We find that $-\langle T_{tt} \rangle + \langle T_{xx} \rangle + \langle T_{yy}\rangle = 0$, which states that the stress energy tensor is indeed traceless (when in Lorentzian signature), as it should be for a conformally invariant theory on the boundary. We also find that $\Omega = - \langle T_{yy} \rangle$. In the normal phase case\footnote{Owing to the simplification in functions, normal phase and the thermodynamic quantities are all determined by the parameter $\phi_{h1} = - \mu/z_{h}$. Hence, selecting a range of values of $\phi_{h1}$ or $\mu$ lets us acquire $\Omega$.}, see \eqref{eqn:reiss_nord_sol}, we must have $h(z) = 1$ and as such $h_{b3}$ becomes zero (since $h_{b0} = 1$ automatically) while function $f(z)$ identifies the $z^3$ coefficient as $f_{b3} = - \left(\frac{1}{z_{h}^3} + \frac{\mu^2}{8 z_{h}} \right)$, when setting $L=1$. Additionally, $\phi_{h1} = -\mu/z_{h}$ by virtue of its functional solution $\phi(z) = \mu(1 - \frac{z}{z_{h}})$. The entropy calculated is the Bekenstein-Hawking entropy, $S$, given by the formula 
\begin{equation}
\label{eqn:eb_entropy_formula}
S = \frac{2 \pi A_{h}}{\kappa_{(4)}^{2}} = \frac{2 \pi V}{z_{h}^2 \kappa_{(4)}^{2}}\,.
\end{equation}
To produce meaningful values of both $S$ and $\Omega$, we make the quantities dimensionless by introducing the necessary factors of $\mu$.
Plots of the grand potential and entropy for both normal (blue curve) and condensing (red, dashed curve) phases at each $g_{\text{YM}}$ are presented in Figures \ref{fig:gcp1plot} to \ref{fig:ent4plot}. Here we confirm the nature of the phase transitions at each $g_{\text{YM}}$. Figures \ref{fig:gcp1plot} and \ref{fig:ent1plot} detail $\Omega$ and $S$ for $g_{\text{YM}} = 3/5$ and show different behaviour to the other $g_{\text{YM}}$. Starting with $\Omega$ at higher temperatures (far right side of the plot), the blue normal curve starts as the only solution, until approximately $T/\mu \approx 0.0059$ after which the condensed phase curve emerges, and exhibits a two-branch ``swallow tail" form. The blue curve continues to be the smaller-valued, preferred $\Omega$ until a critical temperature where it intersects the red condensed phase curve, after which the condensed curve becomes preferable. This intersection is continuous, but not differentiable, which indicates a first order phase transition. This outcome is further confirmed by the entropy plot in Figure \ref{fig:ent1plot}. At $T=T_{c}$ clearly the entropy is not continuous between normal and condensed phases, as there is a jump between the blue curve and lowest red curve branch. As for the larger $g_{\text{YM}}$ values we chose to study, they all display second order phase transition behaviour. In Figures \ref{fig:gcp2plot}, \ref{fig:gcp3plot} and \ref{fig:gcp4plot}, $\Omega$ is both continuous and differentiable, while $S$ is continuous but not differentiable at $T= T_{c}$: the defining characteristics of second order phase transitions. 

These results are in keeping with the previous literature, in particular Figures \ref{fig:gcp4plot} and \ref{fig:ent4plot} show similar behaviour to the results of \cite{Arias:2012py} (see their Figures 4 and 5)\footnote{upon redefining the temperature by a factor of 1/2.}. As for the identification of a critical $g_{\text{YM}}$, \cite{Ammon:2009xh} initially showed that there exists a critical coupling value ``$\alpha_{c}$" (analogous to the reciprocal of $g_{\text{YM}}$ in our work) in 5-dimensions where above $\alpha_{c}$, one finds first order transitions and below they are second order. Similar results were found in \cite{Cai:2013aca} which used two parameters: scalar field mass and charge\footnote{To our knowledge, the specific result of a critical $g_{\text{YM}}$ in four dimensions is new.}.

\clearpage

\section{Black hole interior}
\label{s:int_behaviour}
By virtue of the numerical solutions to the equations of motion, we can study the black hole interior by extending the bulk radial coordinate range from $z=z_{h} =1$ toward $z \to \infty$. The black hole interior has previously been investigated, for example \cite{Donets:1996ja,Breitenlohner:1997hm}. However, the emergence of Kasner geometry in the interior, in the context of holographic superconductors, was initially identified in \cite{Frenkel:2020ysx,Hartnoll:2020rwq,Hartnoll:2020fhc}, where the typical scalar field condensation was utilised. Following this, the interior of vector condensate holographic superconductors appeared in \cite{Cai:2021obq} where numerous Kasner universes were found. The analysis of the interior presented in this section further explores the $\omega_{h0}$ parameter space. The $\omega_{h0}$ values correspond to the coefficient of the horizon expansion in \eqref{eqn:model_horizon_series}. Note that $\omega_{h0}$ is only a function of the temperature\footnote{$\omega_{h0}$ is the last free horizon parameter once the boundary conditions are used. In this sense, it can be thought of as defining $h_{h0}$ and $\phi_{h1}$ which in turn, define $T/\mu$. In reference to section \ref{s:model} we can therefore see our parameters as either ($T/\mu, g_{\text{YM}}$) or ($\omega_{h0}, g_{\text{YM}}$). We chose $\omega_{h0}$ instead of temperature out of numerical convenience.}.

\subsection{Josephson oscillations}
\label{ss:int_field_behaviour}

\begin{figure}[htb!]
\begin{center}
\includegraphics[width=15cm, height=5.5cm]{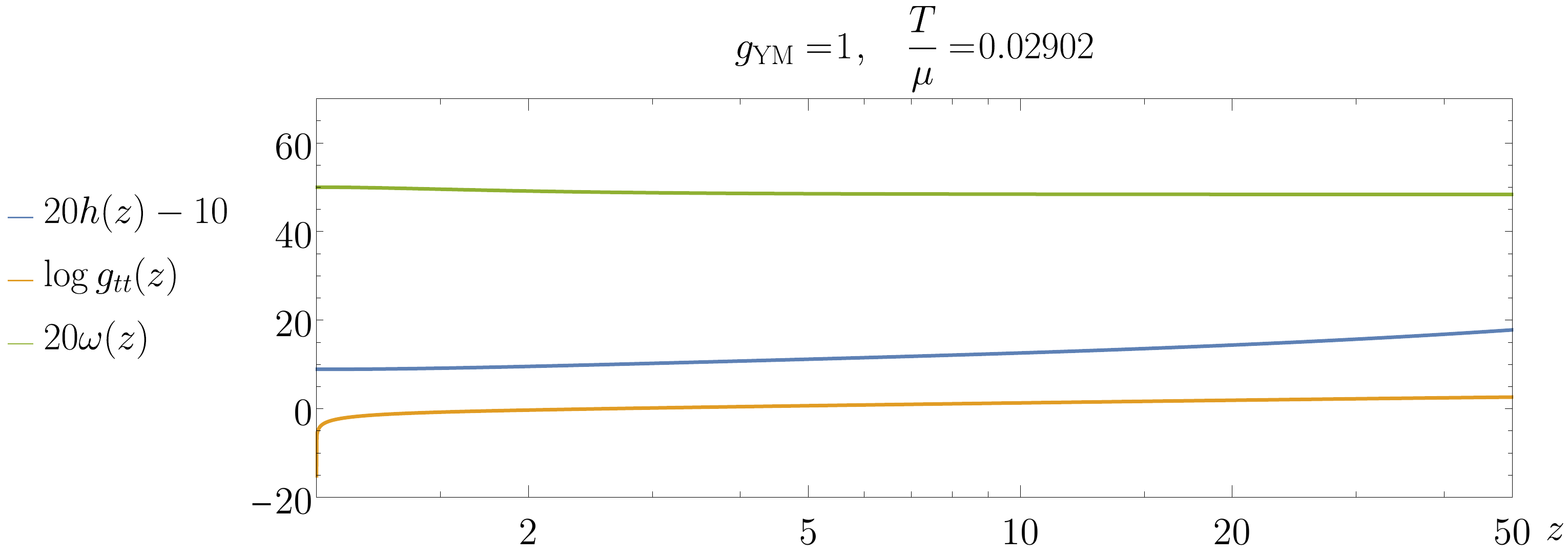}
\end{center}
\caption{Low temperature plot of the black hole interior functions. The functions plotted $h(z)$, $\omega(z)$, $\log{(g_{tt})}$ exhibit tame behaviour in the early interior which is generally the case for lower temperatures.}
\label{fig:intplot_lowtemp}
\end{figure}

\begin{figure}[htb!]
\begin{center}
\includegraphics[width=15cm, height=6cm]{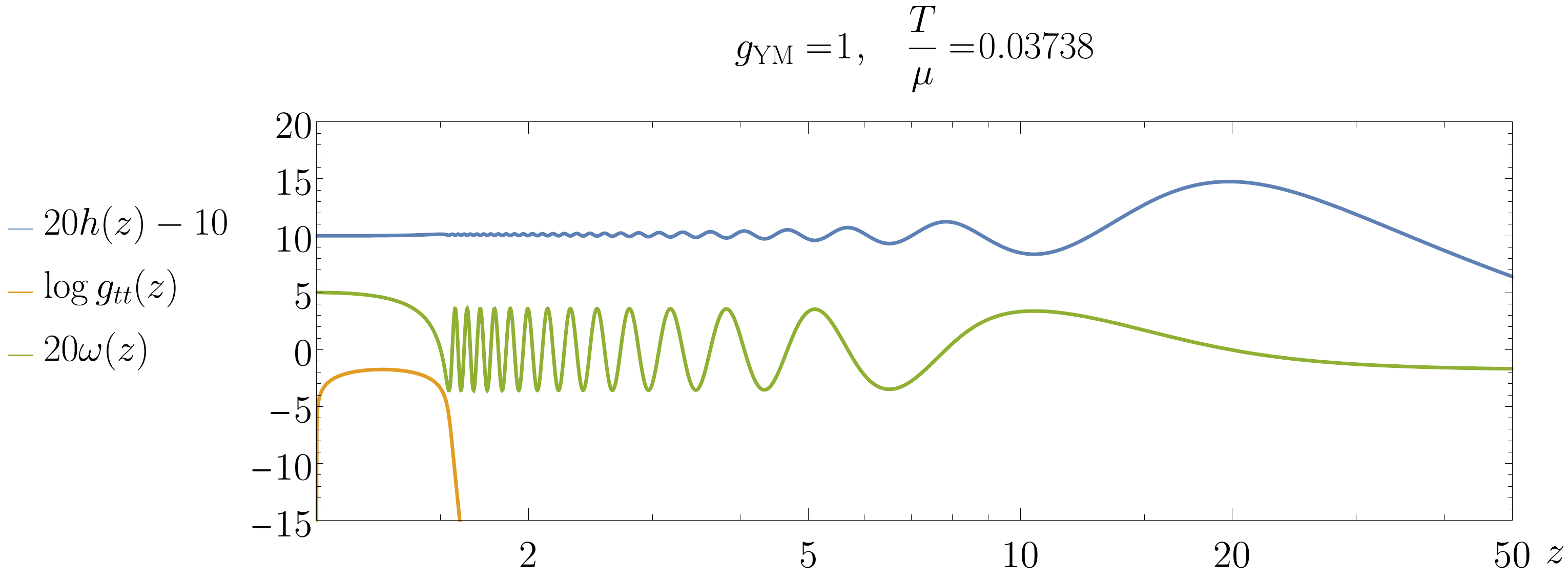}
\end{center}
\caption{High temperature plot of the black hole interior functions. Far more interesting phenomena emerge for functions $h(z)$, $\omega(z)$, $\log{(g_{tt})}$, when close to critical temperature. Here we identify the Josephson oscillations of condensate function $\omega(z)$, while the large decrease in $\log{(g_{tt})}$ at $z \approx 1.5$ is known as the collapse of the Einstein-Rosen bridge. }
\label{fig:intplot_hightemp}
\end{figure}

We begin by providing some typical interior plots which demonstrate the common phenomena. Figures~\ref{fig:intplot_lowtemp} and \ref{fig:intplot_hightemp} introduce an initial view of the interior by plotting particular metric and condensate functions. Both plots extend over a small radial range $z \in (1, 50]$ and take $g_{\text{YM}} = 1$. Figure \ref{fig:intplot_lowtemp} depicts a low temperature $T/\mu = 0.02902$ where the metric functions take rather simple form. On the other hand, pushing closer to critical temperature, $T_{c}/\mu = 0.03748$, far more interesting phenomena emerge in Figure \ref{fig:intplot_hightemp} for $T/\mu = 0.03738$. Firstly $\log{g_{tt}}$, begins to rapidly decline around $z \approx 1.5$ which is directly followed by rapid oscillations in $\omega(z)$ and $h(z)$. These are the Einstein-Rosen (ER) bridge collapse and Josephson oscillations\footnote{More specifically the Josephson oscillations are associated to the oscillations in $\omega(z)$ rather than $h(z)$, as this is the condensate's dual field.}. Notably, the presence of the ER bridge collapse produces these first oscillations, and as will be seen shortly, further changes in the $\log{g_{tt}}$ function also indicate the presence of different Kasner universes. Interestingly, $h(z)$ has twice the frequency of $\omega(z)$. The reason for this can be traced back to analytic results coming from the simplified equations of motion \cite{Cai:2021obq}.

\FloatBarrier
\subsection{Kasner regime}
\label{ss:simplified_eom}
To discover the Kasner universes, we must explore a larger $z$-coordinate range. Just why these Kasner geometries appear in this regime can be understood from the full numerical solutions as well as from a simplified set of analytic solutions. These solutions are given below and the analysis of the interior and graphical representation of the alternations follows. 

\clearpage

\begin{figure}[htb!]
\begin{center}
\includegraphics[width=15cm, height=6cm]{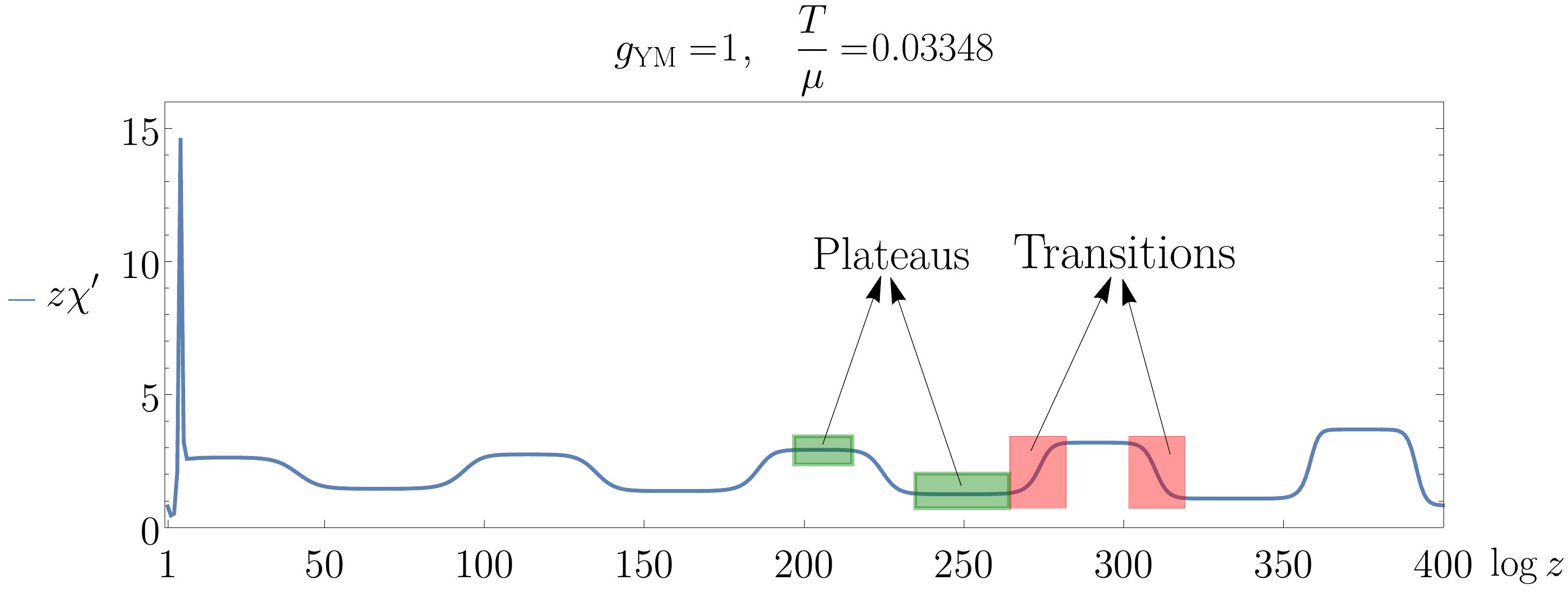}
\end{center}
\caption{An example of the interior of the black hole demonstrating function $z \chi'$ for large $z$, at $g_{\text{YM}} = 1$ and $T/\mu = 0.03348$. On this data we see that up to $\log{z} = 400$, there are five oscillations in $z \chi'$. The different plateaus (two highlighted in green) of each oscillation correspond to a different $n_{h}$ and thus different Kasner universes. The transitions (two highlighted in red) detail how neighbouring Kasner universes connect. }
\label{fig:plat_tran_fig}
\end{figure}

For the current model, by removing the terms which are negligible we arrive at the set of simplified EOM
\begin{align}
\label{eqn:ib_sols_of_simp_eoms}
f' &= \frac{3 f}{z} + \frac{z f (h')^2}{h^2} \nonumber\\
\chi' &= \frac{2 z (h')^2}{h^2} \nonumber\\
h'' & = \frac{2 h'}{z} - \frac{f' h'}{f} + \frac{(h')^2}{h} +\frac{h' \chi'}{2} \\
\phi'' &= -\frac{\phi' \chi'}{2} \nonumber\\
\omega'' &= -\frac{f' \omega'}{f} + \frac{2 h' \omega'}{h} + \frac{\chi' \omega'}{2} \nonumber
\end{align}
which have the resulting analytic solutions
\begin{gather}
\begin{gathered}
\label{eqn:ib_function_solutions_in_large_z}
f(z) \approx f_{0} z^{3 + n_{h}^2}\,, \quad \chi(z) \approx 2 n_{h}^2 \log(z) + \chi_{0}\,, \quad h(z) \approx h_{0} z^{n_{h}}\,, \\
\quad \omega(z) \approx \frac{z^{2(n_{h}-1)}}{2(n_{h}-1)}\omega_{1} + \omega_{0}\,, \quad \phi(z) \approx \frac{z^{1- n_{h}^2}}{1-n_{h}^2}\phi_{1} + \phi_{0}\,,
\end{gathered}
\end{gather} 
where $f_{0}$, $\chi_{0}$, $h_{0}$, $\omega_{1}$, $\omega_{0}$, $\phi_{1}, \phi_{0}$ and $n_{h}$ are all constants. The simplicity of these solutions reveals the Kasner geometry previous mentioned. By taking proper time to be $\tau =  \tau_{0} z^{-\frac{1}{2}(3 + n_{h}^2)}$, along with the simplified solutions \eqref{eqn:ib_function_solutions_in_large_z}, the metric adopts the following form
\begin{equation}
\label{eqn:ib_kas_metric}
ds^2 = -d\tau^2 + c_{t} \tau^{2 p_{t}} dt^2 + c_{x} \tau^{2 p_{x}} dx^2 + c_{y} \tau^{2 p_{y}} dy^2\,,
\end{equation} 
where $\tau_{0}$, $c_{t}$, $c_{x}$, $c_{y}$, $p_{t}$, $p_{x}$ and $p_{y}$ are constants\footnote{Suitable choice of $\tau_{0}$ allows one to scale the $d\tau^2$ coefficient to -1.}. This is a Kasner universe and the connection between the Kasner exponents, $p_{t}$, $p_{x}$, $p_{y}$, and $n_{h}$ is:
\begin{equation}
\label{eqn:ib_kas_exponents_and_nh}
p_{t} = \frac{(n_{h}^2 - 1)}{3+ n_{h}^2}, \quad p_{x} = 	\frac{2(1 - n_{h})}{3 + n_{h}^2}, \quad p_{y} =  \frac{2(n_{h} + 1)}{3 + n_{h}^2}\,.
\end{equation}
One can indeed verify these exponents determine a Kasner universe since $\sum_{i=1}^{3} p_{i} =  \sum_{i=1}^{3} p_{i}^2 = 1$ for $i = {1,2,3} = {t,x,y}$. To clearly see the alternations between different Kasner universes, Figure \ref{fig:plat_tran_fig} plots $z \chi'$ from the horizon to a large radial $z$ value. This function presents a way to visually interpret the different Kasner universes that appear, as it either takes on different constant values for a given $z$ which we refer to as ``plateaus" (the blocks in green), or it transitions between them (blocks in red). These plateaus' constant values are $ z \chi' = 2n_{h}^2$, which determines $n_{h}$ and in turn defines the exponents \eqref{eqn:ib_kas_exponents_and_nh}. As will be seen in future figures, the transitions (see \cite{Hartnoll:2020fhc,Dias:2021afz}) inbetween are brought about by stationary points of the $\log{g_{tt}}$ function.

\begin{figure}[h!]
\centering
\begin{subfigure}[b]{.49\linewidth}
\includegraphics[width=1.1\linewidth]{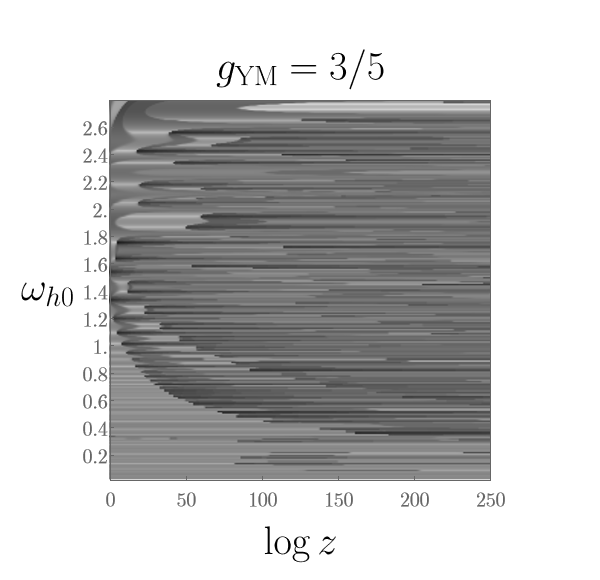}
\label{fig:0.6_plot}
\end{subfigure}
\begin{subfigure}[b]{.49\linewidth}
\includegraphics[width=1.1\linewidth]{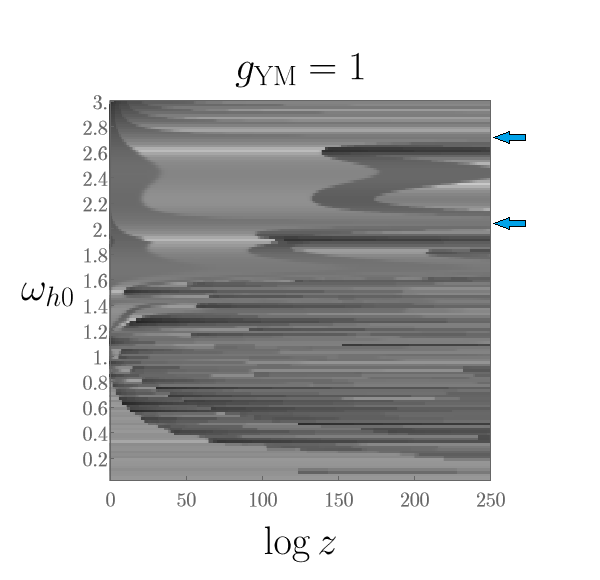}
\label{fig:1_plot}
\end{subfigure}

\begin{subfigure}[b]{.49\linewidth}
\includegraphics[width=1.1\linewidth]{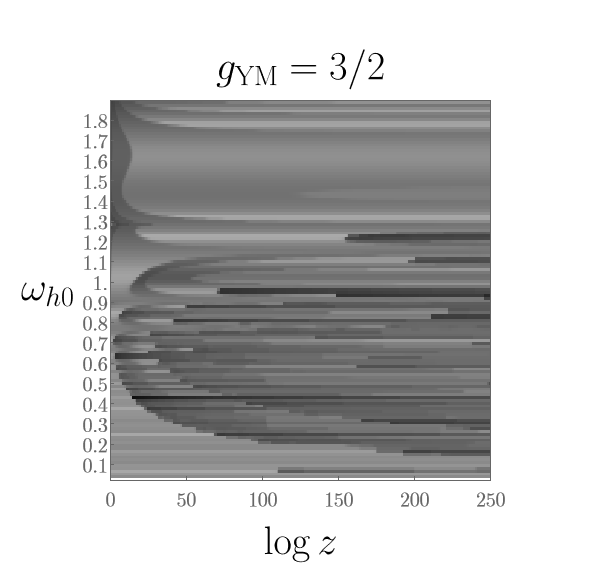}
\label{fig:1.5_plot}
\end{subfigure}
\begin{subfigure}[b]{.49\linewidth}
\includegraphics[width=1.1\linewidth]{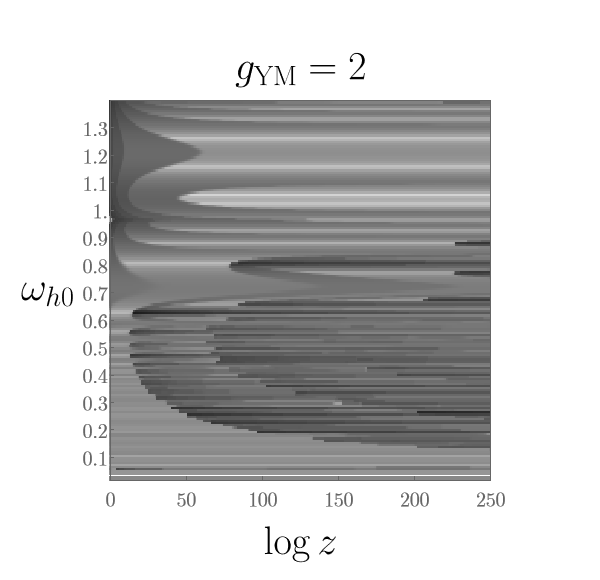}
\label{fig:2_plot}
\end{subfigure}
\caption{Density plots for function $z \chi'$ at the four values of $g_{\text{YM}}$: $3/5, 1, 3/2$ and $2$. The chaotic nature at small $\omega_{h0}$ (and correspondingly, high temperatures) is visible, while more structure arises at larger $\omega_{h0}$ choices. Particular values of $\omega_{h0}$ correspond to critical points, denoted by the blue arrows.}
\label{fig:arrayplots}
\end{figure}

\clearpage

\subsection{Near-oscillatory Kasner epoch}
\label{ss:near_oscillatory_kasner_epoch}

Firstly, to observe the general complicated behaviour in the interior as a function of temperature, Figure \ref{fig:arrayplots} provides density plots of the function $z \chi'$ for bulk radial coordinate $\log{(z)}$ (x-axis) vs. horizon parameter $\omega_{h0}$ (y-axis). Since choice of $\omega_{h0}$ defines $\phi_{h1}$ as well as the UV boundary values ($\mu, \rho$ e.t.c.) for a given solution via our shooting method, then choice of $\omega_{h0}$ essentially sets the temperature $T/\mu$. Hence the parameter spaces of $\omega_{h0}$ and $T/\mu$ are somewhat interchangeable, but we will refer to $\omega_{h0}$ directly in most cases.  The four plots for $g_{\text{YM}} = 3/5$, $1$, $3/2$, $2$ all demonstrate complicated behaviour at small $\omega_{h0}$ (high temperature). For example, for $\omega_{h0} = (0.1, 1.1)$ in the $g_{\text{YM}} = 3/2$ plot, the density changes drastically so the various $n_{h}$ values for each plateau do also. This behaviour spans a wider range of $\omega_{h0}$ in the smaller $g_{\text{YM}}$ value plots\footnote{Note that these plots were generated with an averaging approach due to time constraints on the numeric calculations. In more detail, the numerical integration up to $\log{z} = 250$ for some $\omega_{h0}$ values was unable to complete in reasonable time. When this occurred, the $z \chi'$ value associated to this $\omega_{h0}$ was then filled in by taking an average of the $z\chi'$ values adjacent to it in the $\omega_{h0}$ parameter space.}.

The key result of this paper is that while exploration of the parameter space yields generally complicated functions, there are values of the temperature where the functions become stable and almost oscillatory. To be more specific, the alternations between different Kasner epochs at these particular temperatures become far more regular.

\begin{figure}[htb!]
\begin{center}
\includegraphics[width=15cm, height=6cm]{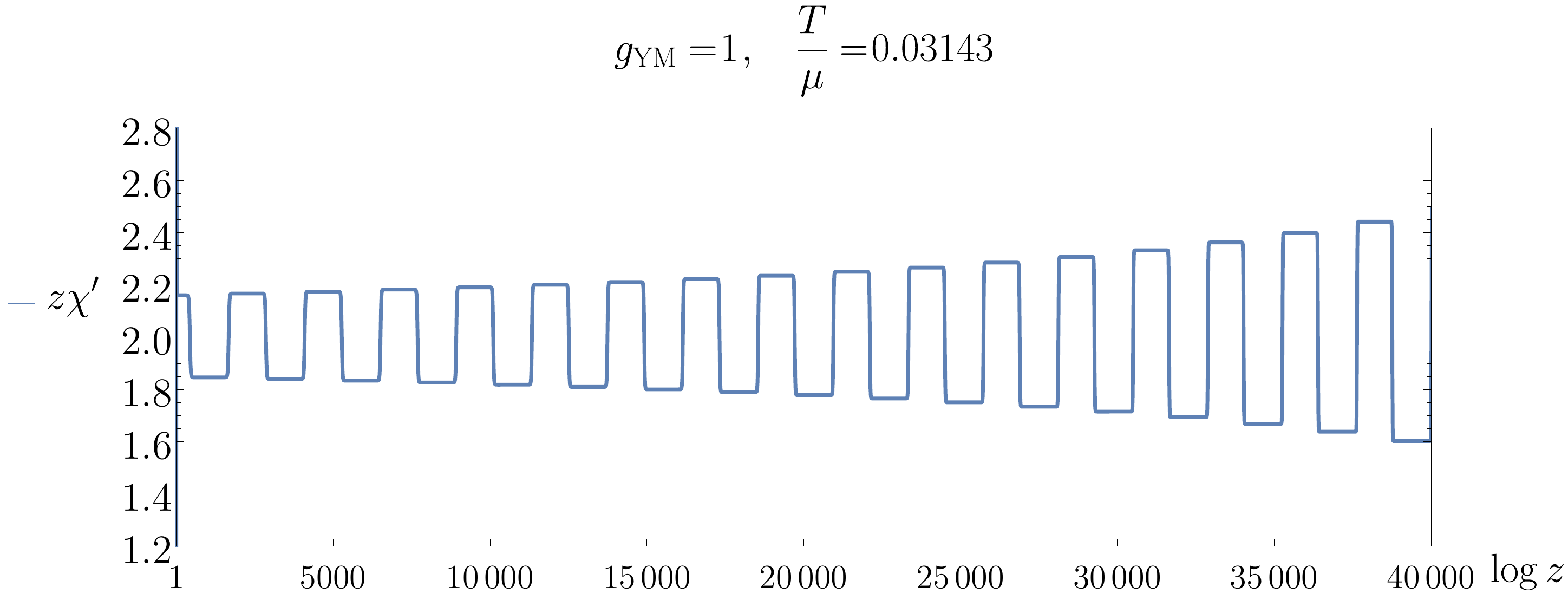}
\end{center}
\caption{Large-$z$ plot for function $ z \chi'$ at $\omega_{h0} = 2.05$. The function demonstrates a large number of oscillations by virtue of the many plateaus. These plateau oscillations are centred around $z \chi' = 2$, which is the value that sets $n_{h} = 1$. The oscillation amplitude grows with $z$ and at very large $z$ the numerics becomes unstable.}
\label{fig:large_z_plot_omega_star_6}
\end{figure}

\begin{figure}[htb!]
\begin{center}
\includegraphics[width=15cm, height=6cm]{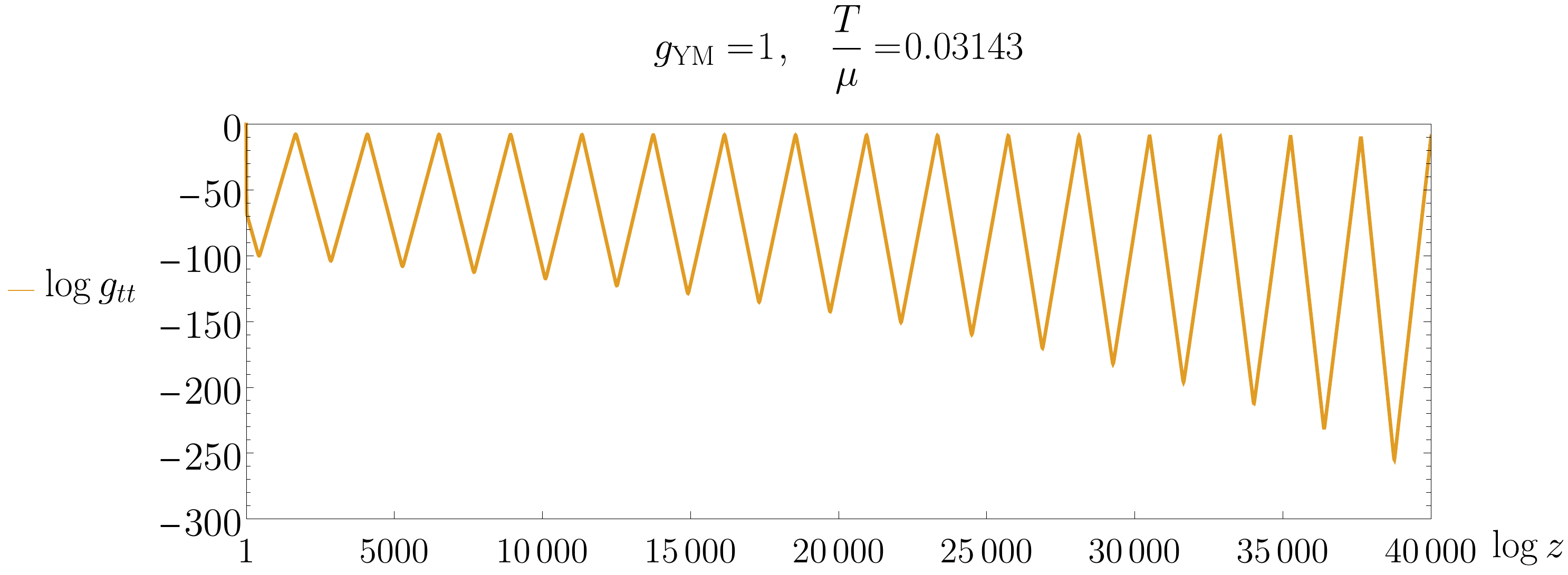}
\end{center}
\caption{Large-$z$ plot for function $\log{g_{tt}}$ at $\omega_{h0} = 2.05$. The function demonstrates a large number of oscillations. The oscillation amplitude grows with $z$. }
\label{fig:large_z_plot_omega_star_5}
\end{figure}

As an example, take $\omega_{h0} = 2.05$ on the $g_{\text{YM}}=1$ density plot of Figure \ref{fig:arrayplots}. A change in density along this value indicates that we may observe many oscillations, hence prompts exploration around this value. The plots of $\log{g_{tt}}$ and $z \chi'$ with large radial range corresponding to this particular $\omega_{h0} = 2.05$ value, are given in Figures \ref{fig:large_z_plot_omega_star_6} and \ref{fig:large_z_plot_omega_star_5}. Here we see the oscillatory nature more clearly, with numerous oscillations extending deep towards the black hole singularity, plotted up to $\log{z} = 40000$. The amplitude of oscillations in both figures increases as we approach larger $z$. Our numerics was shown to breakdown past these large-$z$ values\footnote{We thank Sean Hartnoll for correspondence on this matter.}.  It is a natural question to pose whether fine tuning $\omega_{h0}$ can lead to an infinite number of oscillations that extend all the way to the singularity. The answer to this appears to be that we can approach this highly oscillatory behaviour and at some point the number of oscillations goes to infinity. This is indicated by the lower blue arrow in the $g_{\text{YM}} = 1$ plot, Figure \ref{fig:arrayplots}. We aim to provide evidence of this by observing the relation between the wavelength of $\log{g_{tt}}$ vs. $\omega_{h0}$, as well as the $n_{h}$ Kasner parameter in the subsequent section.

\clearpage

\begin{figure}[h]
    \centering
    \begin{minipage}{0.5\textwidth}
        \centering
        \includegraphics[width=0.9\textwidth]{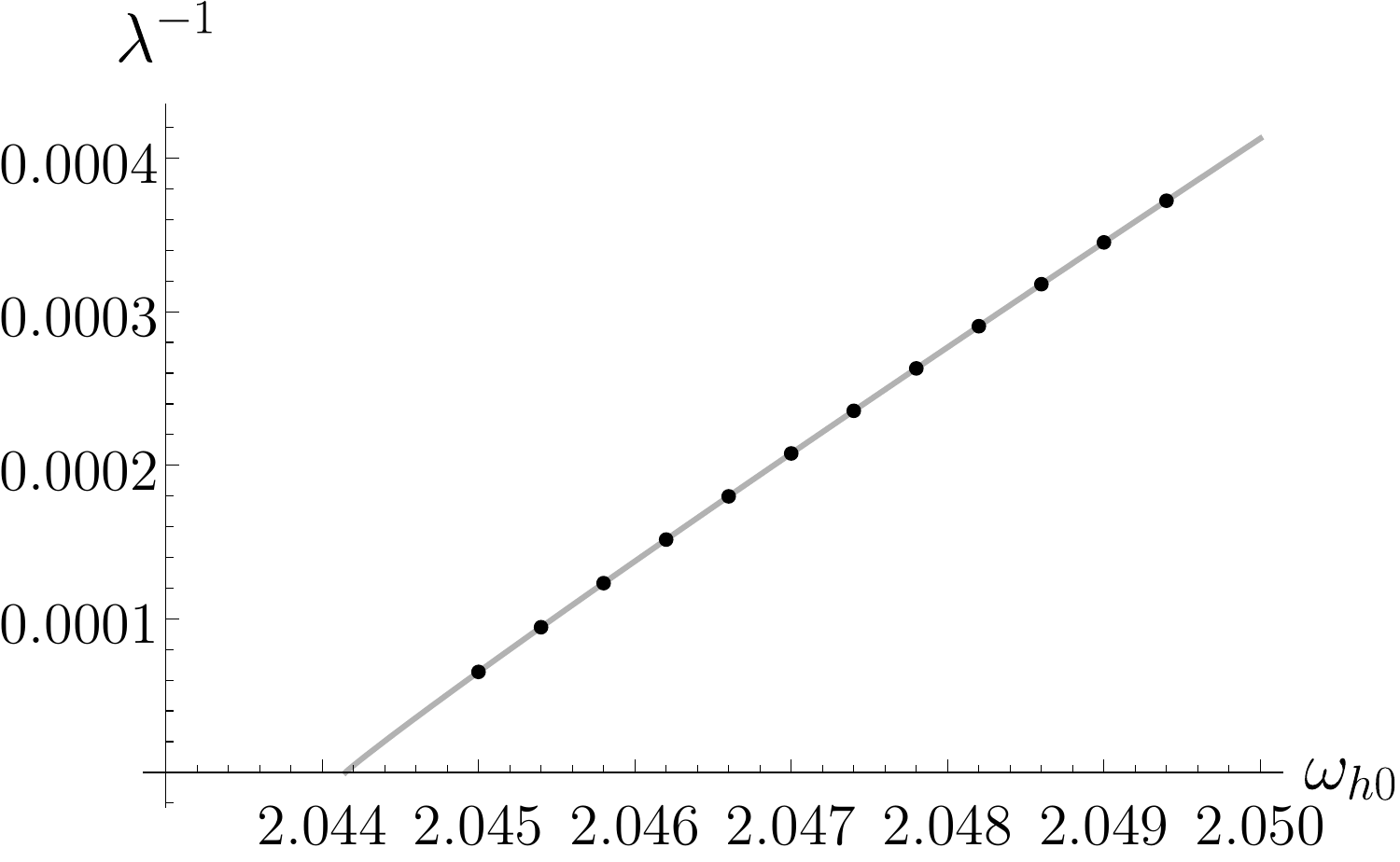} 
    \end{minipage}\hfill
    \begin{minipage}{0.5\textwidth}
        \centering
        \includegraphics[width=0.9\textwidth]{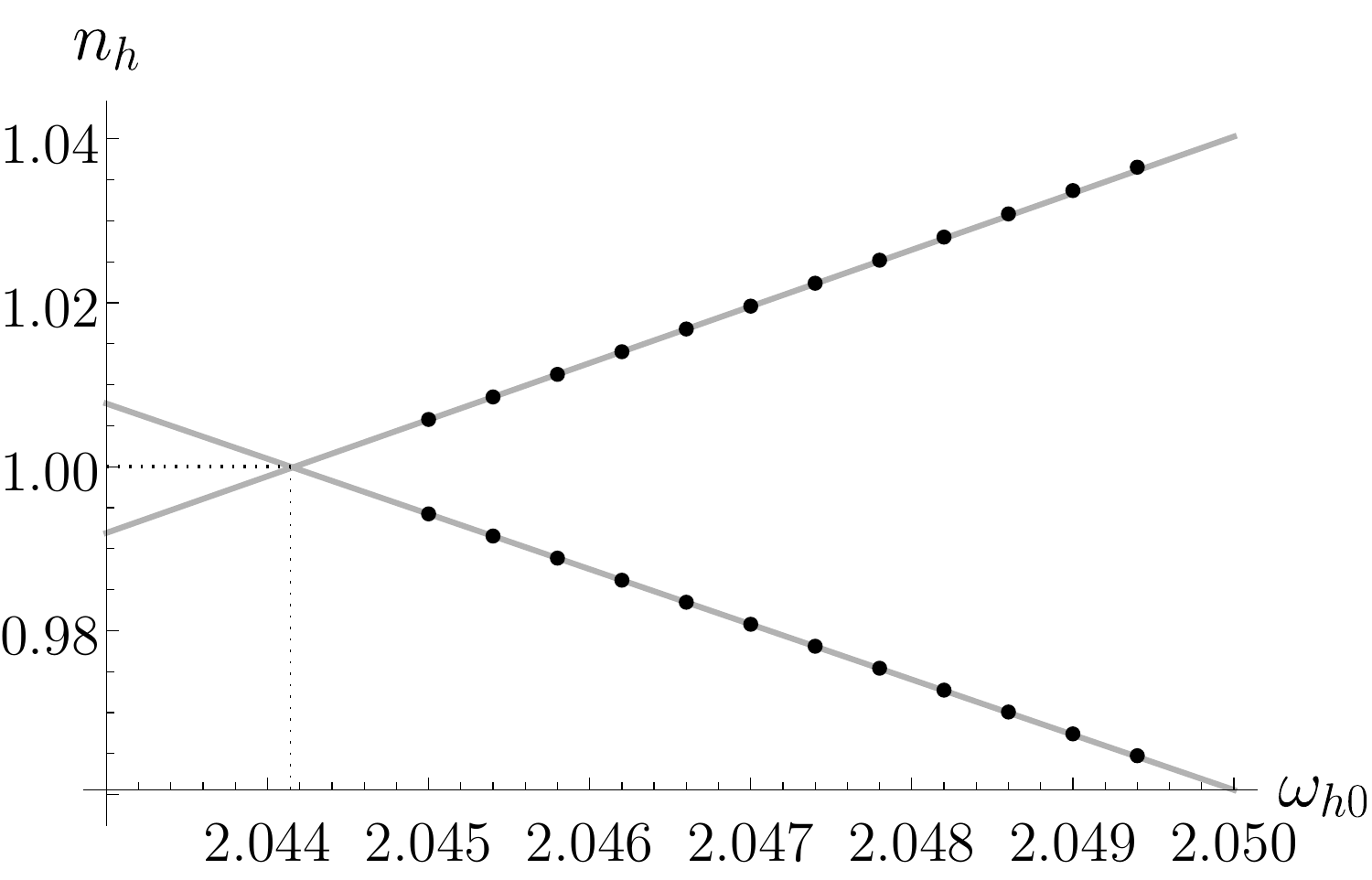}
    \end{minipage}
	\caption{As seen in Figure \ref{fig:large_z_plot_omega_star_5}, $\log{g_{tt}}(z)$ is periodic in $\log{z}$. The left plot demonstrates the reciprocal wavelength of $\log{g_{tt}}$ (labelled $\lambda^{-1}$) as a function of $\omega_{h0}$. The data collected (black points) are well fitted to a simple power law model (grey curve), with exponent $0.96 \pm 0.01$. The right plot displays $n_{h}$, see equation \eqref{eqn:ib_function_solutions_in_large_z}, as a function of $\omega_{h0}$ for the first two plateaus. The data for either of the plateaus are well fitted to a linear function.}
	\label{fig:nh_disp_fig}
\end{figure}

Understanding how varying $\omega_{h0}$ can lead to a large number of Kasner epoch alternations can be achieved as follows. We calculate how the reciprocal wavelength of $\log{g_{tt}}$ (labelled $\lambda^{-1}$) changes upon varying $\omega_{h0}$. The data range explored is from $\omega_{h0} = (2.045, 2.05)$ and the results are presented in left plot of Figure \ref{fig:nh_disp_fig}. As the figure shows, the data are in good agreement with the fitted model
\begin{equation}
\label{eqn:ib_disp_model}
\lambda^{-1} = a (\omega_{h0} - \omega_{*})^{b}
\end{equation}
with 
\begin{equation}
\label{eqn:ib_coefs}
a = 0.056\pm 0.002\,, \qquad b = 0.96 \pm 0.01 \,, \qquad \omega_{*} = 2.04415 \pm 0.00001 \,.
\end{equation}
At $\omega_{h0 } = \omega_{*}$, the wavelength diverges. 
At this point we expect that the first plateau extends to infinity and $\log{g_{tt}} = \text{const.}\,$. The right plot of Figure \ref{fig:nh_disp_fig} displays $n_{h}$ as a function of $\omega_{h0}$. The data are taken from the first two plateaus and a linear fit is applied. The fitted value of $\omega_{*}$ agrees with the value found from the $\lambda^{-1}$ analysis. Similar fits on the other side of the critical point ($\omega_{h0}< \omega_{*}$) match the coefficients of equation \eqref{eqn:ib_coefs}.

The plot also indicates by the dashed curves that both $n_{h} \to 1$ as $\omega_{h0} \to \omega_{*}$. Note that $n_{h} = 1$ corresponds to a special Kasner geometry since from \eqref{eqn:ib_kas_exponents_and_nh} we have $p_{t} = p_{x} = 0$ and $p_{y} = 1$. Plugging these $p_{i}$ into \eqref{eqn:ib_kas_metric} and finally making the coordinate transformation
\begin{equation}
\label{eqn:ib_coord_trans_kas_to_mink}
\hat{\tau} = \sqrt{c_{y}}\tau \cosh{\left(\sqrt{c_{y}} y\right)}\, \quad \hat{y} = \sqrt{c_{y}}\tau \sinh{\left(\sqrt{c_{y}}y \right)} \,, 
\end{equation}
we find that the metric, up to choice of constants $c_{i}$, is flat spacetime
\begin{equation}
\label{eqn:ib_flat_metric}
ds^2 \big|_{n_h = 1}  = - \frac{1}{c_{y}} d\hat{\tau}^2 + c_{t}dt^2 + c_{x} dx^2 + \frac{1}{c_{y}} d\hat{y}^2 \,.
\end{equation}
$\tau$ and $y$ are the well known Rindler coordinates, however, instead of the time and radial $z$-direction, we transform the time and $y$-direction.

\begin{figure}[htb!]
\begin{center}
\includegraphics[width=15cm, height=6cm]{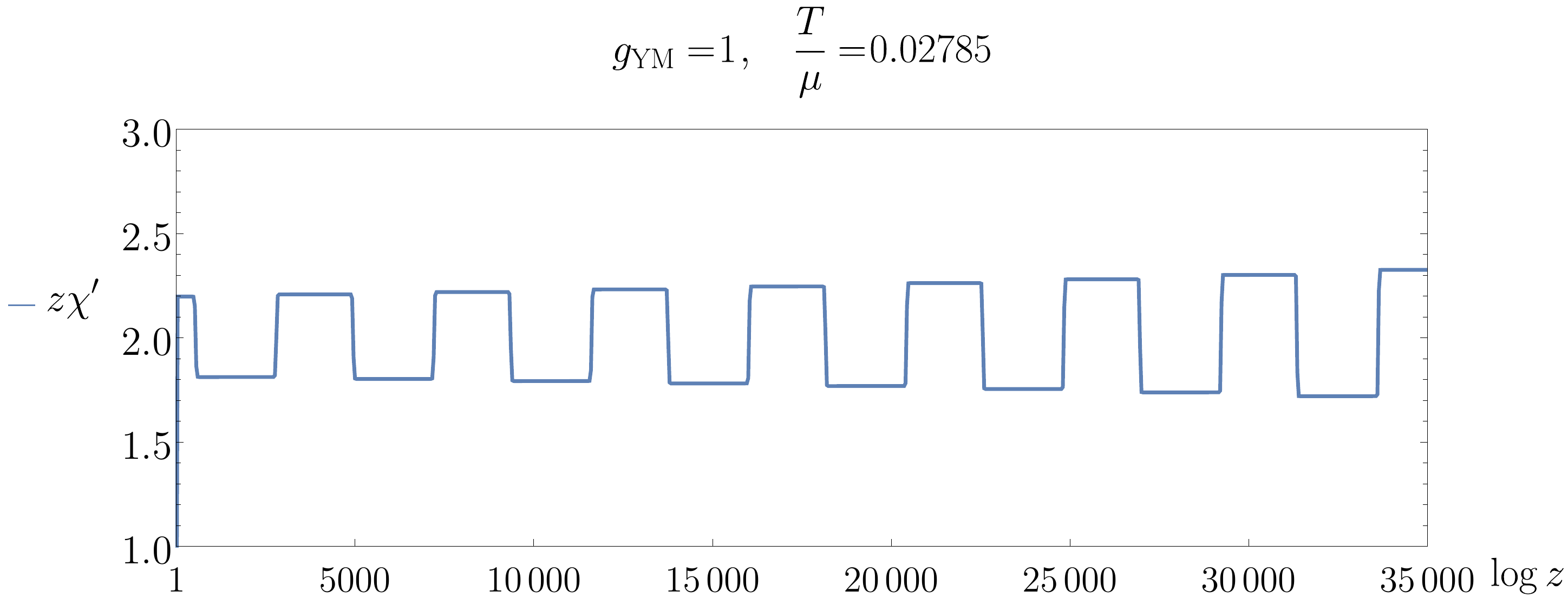}
\end{center}
\caption{Large-$z$ plot for $z \chi'$ at $\omega_{h0} = 2.713$. The function once again demonstrates a large number of oscillations at this $\omega_{h0}$. The function is shown to oscillate around $z \chi' =2 $ or analogously $n_{h} = 1$.}
\label{fig:large_z_plot_omega_star_2713_zchiprime}
\end{figure}

\begin{figure}[htb!]
\begin{center}
\includegraphics[width=15cm, height=6cm]{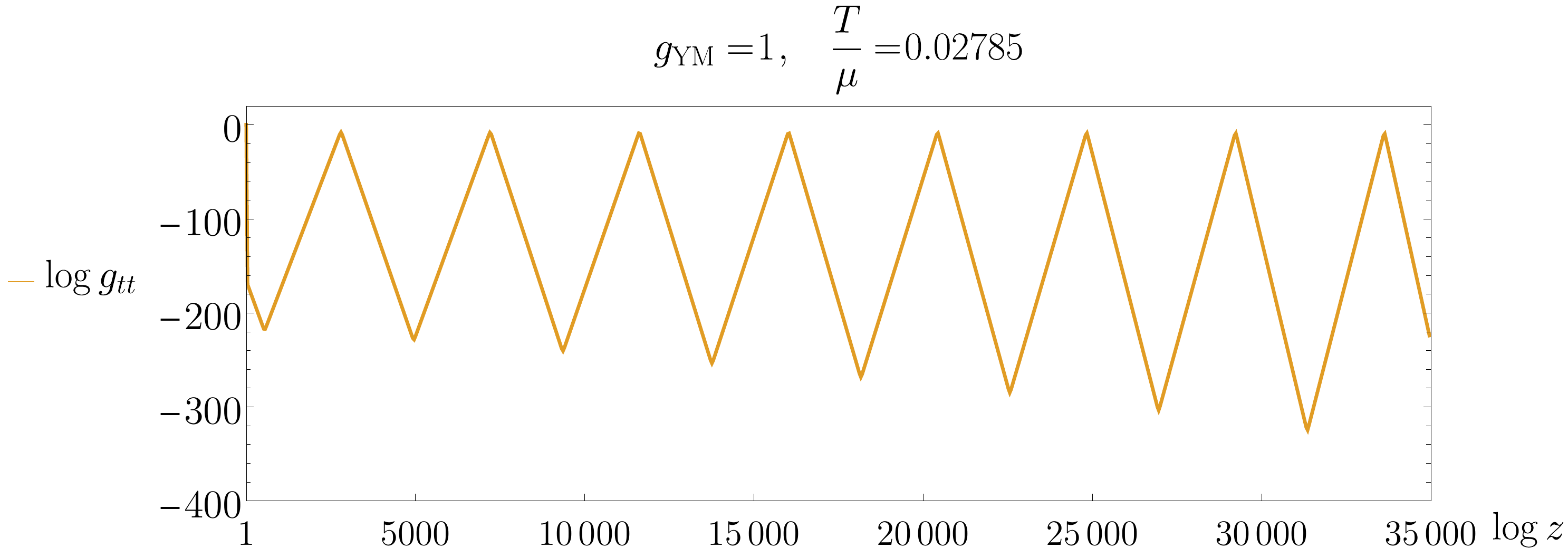}
\end{center}
\caption{Large-$z$ plot for $\log{g_{tt}}$ at $\omega_{h0} = 2.713$. The function demonstrates a large number of oscillations, with increasing amplitude. }
\label{fig:large_z_plot_omega_star_2713_loggtt}
\end{figure}

Ultimately the analysis tells us that at $\omega_{h0} = \omega_{*}$, we obtain flat geometry. Moving slightly away from this value leads to many oscillations between Kasner universes centred around the $n_{h} = 1$ value. The parameter space can be scanned and it is possible to find additional critical points. For example at $g_{\text{YM}}=1$, $\omega_{h0} = 2.713$ the metric functions also exhibit multiple oscillations, see Figures \ref{fig:large_z_plot_omega_star_2713_zchiprime} and \ref{fig:large_z_plot_omega_star_2713_loggtt}. This $\omega_{h0}$ value corresponds to the upper blue arrow in the density plot of Figure \ref{fig:arrayplots}.

 \clearpage

\section{Conclusion}
\label{s:conclusion}

In this paper we used a holographic p-wave superconductor to explore what lies beyond the black hole horizon. Beginning with the black hole exterior, our numerical results showed that phase transitions are dependent on the Yang-Mills coupling parameter $g_{\text{YM}}$ with a critical value that separates first and second order phase transitions. The interior yielded further interesting results upon exploring the parameter space. At specific temperature values the geometry of the interior becomes flat. Approaching these temperature values a large number of Kasner universe alternations appear. It is possible that further investigation of the parameter space could yield other regions of interest. 

\vspace{0.5in}   \centerline{\bf{Acknowledgements}} \vspace{0.2in}

We thank Pau Figueras, Dami\'{a}n Galante  and Sean Hartnoll for useful discussions. LS is supported by an STFC quota  studentship. DV is supported by the STFC Ernest Rutherford grant ST/P004334/1. No new data were generated or analysed during this study.

\printbibliography

\end{document}